\documentclass[12pt]{article}
\usepackage{amsmath,amssymb,latexsym,epsf,epsfig,graphicx,tikz}

\begin{document}

\newcommand{\prt}{\partial}
\newcommand{\II}{\mbox{${\mathbb I}$}}
\newcommand{\CC}{\mbox{${\mathbb C}$}}
\newcommand{\RR}{\mbox{${\mathbb R}$}}
\newcommand{\QQ}{\mbox{${\mathbb Q}$}}
\newcommand{\HH}{\mbox{${\mathbb H}$}}
\newcommand{\ZZ}{\mbox{${\mathbb Z}$}}
\newcommand{\NN}{\mbox{${\mathbb N}$}}
\newcommand{\DD}{\mbox{${\mathbb D}$}}
\newcommand{\PP}{\mbox{${\mathbb P}$}}
\def\G{\mathbb G}
\def\UU{\mathbb U}
\def\S{\mathbb S}
\def\T{\mathbb T}
\def\KK{\mathbb K}
\def\J{\mathbb J}
\def\tT{\widetilde{\mathbb T}}
\def\HH{\mathbb H}
\def\tS{\widetilde{\mathbb S}}
\newcommand{\mB}{{\mathbb B}}
\newcommand{\mA}{{\mathbb A}}
\newcommand{\mC}{{\mathbb C}}
\newcommand{\mM}{{\mathbb M}}
\newcommand{\ma}{{\mathbb a}}

\def\V{\mathbb V}
\def\tV{\widetilde{\mathbb V}}
\newcommand{\D}{{\mathbb D}}
\def\hint{H_{\rm int}}
\def\R{{\cal R}}

\newcommand{\rd}{{\rm d}}
\newcommand{\diag}{{\rm diag}}
\newcommand{\cU}{{\cal U}}
\newcommand{\K}{{\mathcal K}}
\newcommand{\cP}{{\cal P}}
\newcommand{\mT}{{\mathbb T}}
\newcommand{\dQ}{{\dot Q}}
\newcommand{\dS}{{\dot S}}
\newcommand{\dW}{{\dot W}}
\newcommand{\W}{{\mathcal W}}
\newcommand{\Q}{{\mathbb Q}}

\newcommand{\pnf}{P^N_{\rm f}}
\newcommand{\pnb}{P^N_{\rm b}}
\newcommand{\hnf}{P^Q_{\rm f}}
\newcommand{\hnb}{P^Q_{\rm b}}

\newcommand{\ph}{\varphi}
\newcommand{\phd}{\widetilde{\varphi}} 
\newcommand{\jt}{\widetilde{j}}
\newcommand{\phs}{\varphi^{(s)}}
\newcommand{\phb}{\varphi^{(b)}}
\newcommand{\phds}{\widetilde{\varphi}^{(s)}}
\newcommand{\phdb}{\widetilde{\varphi}^{(b)}}
\newcommand{\vt}{\vartheta}
\newcommand{\lambdad}{\widetilde{\lambda}}
\newcommand{\mut}{\widetilde{\mu}}
\newcommand{\nut}{\widetilde{\nu}}
\newcommand{\tx}{\widetilde{x}} 
\newcommand{\td}{\widetilde{d}} 
\newcommand{\etat}{\widetilde{\eta}}
\newcommand{\phl}{\varphi_{{}_L}}
\newcommand{\phr}{\varphi_{{}_R}}
\newcommand{\phz}{\varphi_{Z}}
\newcommand{\mum}{\mu_{{}_-}}
\newcommand{\mup}{\mu_{{}_+}}
\newcommand{\mupm}{\mu_{{}_\pm}}
\newcommand{\muv}{\mu_{{}_V}}
\newcommand{\mua}{\mu_{{}_A}}
\newcommand{\wt}{\hat{t}}

\def\a{\alpha}
 
\def\A{\mathcal A} 
\def\B{\mathcal B} 
\def\H{\mathcal H} 
\def\U{\mathcal U} 
\def\E{\mathcal E} 
\def\C{\mathcal C} 
\def\L{\mathcal L} 
\def\M{\mathcal M} 
\def\O{\mathcal O}
\def\I{\mathcal I}
\def\Z{\mathcal Z} 
\def\der{\partial }
\def\mis{{\frac{\rd k}{2\pi} }}
\def\ri{{\rm i}}
\def\p{{\bf p}}
\def\xt{{\widetilde x}}
\def\ft{{\widetilde f}}
\def\gt{{\widetilde g}}
\def\qt{{\widetilde q}}
\def\tt{{\widetilde t}}
\def\tmu{{\widetilde \mu}}
\def\prt{{\partial}}
\def\tr{{\rm Tr}}
\def\inc{{\rm in}}
\def\out{{\rm out}}
\def\Li{{\rm Li}}
\def\e{{\rm e}}
\def\eps{\varepsilon}
\def\k{\kappa}
\def\v{{\bf v}}
\def\ebf{{\bf e}}
\def\abf{{\bf A}}
\def\lb{\Omega_{{}_{\rm LB}}}
\def\rlb{\rangle_{{}_{\rm LB}}}
\def\hlb{\H_{{}_{\rm LB}}}
\def\llb{\L_{{}_{\rm LB}}}

\def\Om{\Omega_{\rm mix}}

\def\be{\begin{equation}}
\def\ee{\end{equation}}

\def\bea{\begin{eqnarray}}
\def\eea{\end{eqnarray}}

%%%%%%%%%%%%%%%%%%% INIZIO %%%%%%%%%%%%%%%%%%%%%%%

%%%%%%%% 
\newcommand{\finprf}{\null \hfill {\rule{5pt}{5pt}}\\[2.1ex]\indent}

%%%%%%%%%%%%%%%%%%%%%%%
\pagestyle{empty}
\rightline{March 2022}
%\rightline{Preliminary Draft}

\bigskip 

\begin{center}
{\Large\bf Quantum states from mixtures\\ of equilibrium distributions}
\\[2.1em]

\bigskip

{\large Mihail Mintchev}\\ 
\medskip 
{\it  
Istituto Nazionale di Fisica Nucleare and Dipartimento di Fisica, Universit\`a di
Pisa,\\ Largo Pontecorvo 3, 56127 Pisa, Italy}
\bigskip 

\bigskip 
\bigskip 
\bigskip 

\end{center}

%\date{\today}

\begin{abstract}

\bigskip 

We construct and explore a family of states for quantum systems in contact with two or more 
heath reservoirs. The reservoirs are described by equilibrium distributions. The interaction of 
each reservoir with the bulk of the system is encoded in a probability, which characterises the 
particle exchange among them and 
depends in general on the particle momentum. The convex combination of the reservoir distributions, weighted with 
the aforementioned probabilities, defines a new distribution. We establish the existence of an 
emission-absorption regime in which the new distribution generates a non-equilibrium quantum state. 
We develop a systematic field theory framework for constructing this state and illustrate its  
physical properties on a simple model. In this context we derive the particle current full counting 
statistics, the heat current and the Lorenz number. The entropy production and the relative 
quantum fluctuations are also determined.

\end{abstract}

\vfill
%\rightline{IFUP-TH 1/2016}
%\rightline{LAPTH-069/15}
\newpage
\pagestyle{plain}
\setcounter{page}{1}

\section{Introduction}

The study of quantum systems away from equilibrium is an active area of research for many decades.   
A specific feature of the subject is the existence of a wide 
class of non-equilibrium states, which reflects the large number of possible ways to drive 
a system away from equilibrium. Steady states \cite{James-59}-\cite{Kita-10} represent among them a relevant 
subfamily. In the quantum field theory context the mean value of 
the physical observables (densities, currents, entropy production,...) 
in such states is usually time independent, the time dependence showing up in the quantum fluctuations. 
This feature of steady states simplifies the analysis and allows for some degree of explicit and exact treatment. 
It is essential also for the applications to the charge and heat transport in mesoscopic systems (see e.g. \cite{Datta1, Datta2}), 
which attracts recently much attention. 

The complete understanding and systematic description 
of non-equilibrium steady states of quantum systems is a challenging open problem. 
We deal below with a particular aspect of this 
problem, regarding the construction \cite{Ruelle-00, Sasa-06} of probability distributions describing such states. We focus on 
quantum systems which are driven away from equilibrium 
by the contact with two or more heat reservoirs $\{R_i\, :\, i=2,...,N\}$ at different (inverse) temperatures 
$\beta_i$ and chemical potentials $\mu_i$. We assume that the capacity of the reservoirs 
is large enough, so that the process of emission and absorption of particles does not modify 
the values of $\beta_i$ and $\mu_i$. Denoting by $k$ the particle momentum, we associate 
with each reservoir an equilibrium particle distribution 
$d_i(k)$ and a probability $w_i(k)$ of particle exchange with the system, such that  
\be 
\sum_{i=1}^N w_i (k) = 1 \, , \qquad w_i (k) \geq 0\, .
\label{w1}
\ee
With these data one can build the convex combination 
\be 
d_{\rm mix}(k) = \sum_{i=1}^N w_i(k)\, d_i(k)  \, ,
\label{mix1}
\ee
representing the mixed distribution (called also mixture) \cite{mix} of $d_i(k)$ associated to the weights $w_i(k)$. 
Our main goal below is to construct the quantum state $\Omega_{\rm mix}$ generated by the mixture  
(\ref{mix1}) and study its basic features. 
$\Om$ with $k$-independent weights is known to be an equilibrium state. We will show however 
that the equilibrium can be broken if the weights, which physically represent the particle exchange probabilities 
of the system with the reservoirs, depend on the momentum. To characterise the departure from equilibrium, 
we introduce the concept of particle exchange imbalance $I(k)$, which is defined in terms of $w_i(k)$.  
Investigating the dependence of the entropy production on $I(k)$, we 
determine a regime in which $\Om$ is a non-equilibrium state. 

In this paper we focus on reservoirs $R_i$ described by Gibbs ensambles $d_i(k)$ of free 
fermions or bosons. We assume also that the only interaction of the reservoirs with the bulk 
is encoded in the particle exchange probabilities $w_i(k)$. At this point the 
necessary step in applying the mixed distribution (\ref{mix1}) 
in the general context of canonical quantum field theory 
is to construct a representation $\pi_{\rm mix}$ of the algebra of canonical (anti)commutation 
relations $\A$, which is generated by the state $\Om$.   
In fact, by means of $\pi_{\rm mix}$ one can 
build free bulk quantum fields with relativistic or non-relativistic dispersion relations. 
These fields describe the physics of the heat reservoirs and their contact with the system 
and one is left with the choice of bulk dynamics. To start with, we investigate in this article  
two cases: free evolution in the bulk and interaction with an external potential. 
We observe however that in analogy with conventional finite temperature field theory 
\cite{K} in the Gibbs state, one can explore via perturbation theory 
any renormalizable bulk interaction in the state $\Om$. 

The paper is organised as follows. In the next section we derive in explicit form the 
representation $\pi_{\rm mix}$ of the canonical (anti)commutation relation algebra $\A$, 
which is generated by the mixture (\ref{mix1}) of two Gibbs distributions. In section 3 we 
test the physical properties of this representation, exploring the free Schr\"odiger dynamics on the line. 
We determine the mean particle density in the state $\Om$ and  
the full counting statistics associated to the particle current. We explore afterwards 
the heat transport, deriving the Lorenz number and discussing the Wiedemann-Franz law. 
The non-equilibrium features of $\pi_{\rm mix}$ are characterised by means of the particle 
exchange imbalance $I(k)$. We determine a range of the parameters for which $\Om$ breaks time 
reversal invariance and investigate the behaviour of the 
entropy production operator $\dS(t)$ in this regime. We 
prove that $I(k)>0$ implies that the mean value and all zero frequency quantum 
fluctuations of $\dS(t)$ are positive. This statement is interpreted as  a quantum 
version of the second law of thermodynamics in our setup. In section 3 
we briefly describe also the linear response theory to an external potential in the 
representation $\pi_{\rm mix}$. Section 4 contains a brief summary of the results and 
collects the conclusions. Some challenging further developments are also indicated.

\bigskip 

\section{Mixture of two Gibbs distributions}

\begin{figure}[h]
\begin{center}
\begin{picture}(620,40)(80,340) 
\includegraphics[scale=0.95]{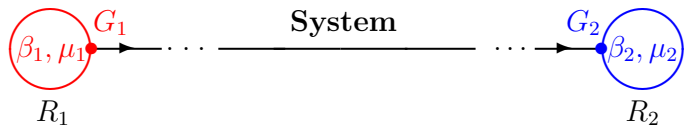}
\end{picture} 
\end{center}
\caption{System connected to two heath reservoirs.} 
\label{fig1}
\end{figure}

In this paper we study systems in one space dimension, which are in contact via the gates 
$G_{1,2}$ with two heat reservoirs $R_{1,2}$ as shown in Fig. \ref{fig1}.
Settings of this type are intensively studies both theoretically and experimentally. 
They are applied for describing the charge and heat transport in quantum 
wires \cite{kf-92}-\cite{KGA-20} and in quantum Hall samples \cite{AS-08}. 
The exciting possibility to realise in laboratory the system in Fig.\ref{fig1} by means of 
ultracold Bose gases \cite{BDZ-08}-\cite{CS-16} is also worth mentioning. 
Such experiments are usually characterised by a remarkable precision, 
which allows to investigate a variety of fundamental aspects of many-body quantum physics. 
Finally, in the context of quantum thermodynamics \cite{qt} the setup in Fig. \ref{fig1} schematically 
represents a quantum heat engine \cite{he} operating between the ``hot" and ``cold" reservoirs $R_1$ 
and $R_2$. 
\bigskip

\subsection{Components of the mixture}

We consider the system in Fig. \ref{fig1} and assume that the gates $G_{1,2}$ are located 
at $\pm \infty$ and the reservoirs $R_{1,2}$ are described by the Gibbs distributions 
\begin{equation}
d^\pm_i(k) = \frac{1}{\e^{\beta_i [\omega_i(k) - \mu_i)} \pm 1}\,   
\label{e10}
\end{equation}  
for fermions $(+)$ and bosons $(-)$. Here $k$ is the particle momentum and $\omega_i(k)\geq 0$ is the dispersion 
relation in $R_i$. In a more general situation for systems which have several conserved charges in involution 
one can take instead of (\ref{e10}) a generalised Gibbs ensemble \cite{J1, J2}. The construction below applies 
with slight modifications to this case as well. 

The probabilities of particle exchange of the reservoirs with the system depend in general on $k$ and satisfy 
\be 
w_1(k) + w_2(k)= 1 \, , \qquad w_i(k) \geq 0\, , \quad \forall k \in \RR\, .
\label{w2}
\ee
According to (\ref{mix1}), with the above data we can construct the mixture \cite{mix}
\be 
d^\pm_{\rm mix} (k) = w_1(k)\, d_1^\pm(k) + w_2(k)\, d_2^\pm(\omega(k))  = 
w_1(k)\, d_1^\pm(k) + [1-w_1(k)]\, d_2^\pm(\omega(k))
\label{mix2}
\ee 
of the two Gibbs distributions. 

Before developing the quantum field theory framework for constructing the state $\Om$ 
associated with (\ref{mix2}), it is instructive to provide an intuitive physical picture of the particle exchange between the 
reservoirs and the system. For this purpose we consider the difference 
\be 
I_i(k) \equiv w_i(k)-w_i(-k) \, , \qquad i=1,2 
\label{Imb1}
\ee 
which can be interpreted as emission-absorption imbalance in the gate $G_i$. By construction 
$I_i(k)$ is a real function with values in the interval $[-1\, ,\, 1]$. The emission of particles 
with momentum $k_0$ from the reservoir $R_i$ dominates the absorption if $I_i(k_0) >0$. The absorption is instead  
predominant if $I_i(k_0) <0$. Using that (\ref{w2}) holds for any $k\in \RR$, one has that 
\be 
I_1(k) + I_2(k)= 0\, . 
\label{Imb3}
\ee
Therefore, if for instance the emission of particles with momentum $k_0$ dominates in the gate $G_1$, 
it is the absorption of such particles which dominates in the gate $G_2$. If the total 
bulk particle number of the system is conserved, 
this means that particles with momentum $k_0$ are flowing from $R_1$ to $R_2$, which 
drives the system away from equilibrium. This heuristic description 
is confirmed below in section 3. 

Summarising, because of (\ref{Imb3}) one can use 
\be 
I(k) \equiv I_1(k) = -I_2(k)\, , 
\label{Imb4}
\ee
for characterising the process of particle exchange of the system with the reservoirs. 
In what follows we refer to $I(k)$ as particle exchange imbalance or simply imbalance. 
We will show below that the value of the imbalance controls the departure from equilibrium. 
More precisely, $I(k) \not= 0$ turns out to be a necessary condition for equilibrium breaking. 
Notice in this respect that if the weights $w_i(k)$ are even functions (in particular $k$-independent), one has 
$I(k)=0$, which leads to equilibrium. 

We conclude this subsection providing some explicit examples for the weights $w_i(k)$. 
In the special case 
\be
w_1(k) = \left (\frac{1}{2}+c_+\right )\theta (k) + \left (\frac{1}{2}+c_-\right )\theta(-k)\, , \quad 
w_2(k) = 1- w_1(k)\, , \quad c_\pm \in [-1/2\, ,\, 1/2]\, , 
\label{w6} 
\ee
where $\theta$ is the Heaviside unit step function, one has constant but different (for $c_+\not=c_-$) 
emission and absorption; in fact, 
the imbalance is  
\be 
I(k) = c_+ - c_- \, . 
\label{I3}
\ee 
Notice that the weights (\ref{w6}) are scale invariant. 

For illustrating the particle transport we will adopt below also  
\be
w_1(k) = \frac{1}{2}\left (1+\e^{-\xi k^2}\right )
\theta (k) + \frac{1}{2}\left (1-\e^{-\xi k^2}\right )\theta(-k)\, , 
\quad w_2(k) = 1-w_1(k)\, , 
\quad \xi >0\, , 
\label{w6b} 
\ee
which lead to the $k$-dependent imbalance   
\be 
I(k) = \e^{-\xi k^2} \, . 
\label{I3c}
\ee
 
\bigskip 

\subsection{Field theory framework - the representation $\pi_{\rm mix}$}

We develop here a quantum field theory formalism based on the mixed distribution (\ref{mix2}). 
In order to treat in a unified way both Fermi and Bose statistics, we introduce a canonical algebra 
$\A$, whose generators $\{a(k),\, a^*(k)\, :\, k \in \RR\}$ satisfy 
either anticommutation ($+$ for fermions) or commutation  
($-$ for bosons) relations respectively: 
\bea
&&\left [a(k)\, ,\, a^*(p)\right ]_{{}_\pm} = 2\pi \delta(k-p)\II_{\A}\, , 
\label{e7a}
\\
&&\left [a(k)\, ,\, a(p)\right ]_{{}_\pm} = \left [a^*(k)\, ,\, a^*(p)\right ]_{{}_\pm} =0 \, , 
\label{e7b}
\eea
where $\II_\A$ is the identity element. 
The central problem now is to construct a Hilbert space representation 
$\{\pi_{\rm mix},\, \H_{\rm mix} , \, \Om \}$ of $\A$, such that 
\be
(\Om ,\, a^*(k) a(p) \Om ) \equiv \langle a^*(k) a(p) \rangle_{\rm mix} = 
2 \pi \delta (k-p) d^\pm_{\rm mix} (k) \, , \qquad \qquad \qquad 
\label{rep1}
\ee
where $(\cdot \, ,\, \cdot )$ is the scalar product in $\H_{\rm mix}$. 
In order to simplify the notation, here and in what follows we adopt the physical 
convention and denote the  
image $\pi_{\rm mix}[a(k)]$ acting as a linear operator in the representation space $\H_{\rm mix}$, 
in the same way as the abstract algebraic element $a(k) \in \A$. With this notation  
the operator $a^*(k)$ is the Hermitian conjugate of $a(k)$ in $\H_{\rm mix}$. 

The explicit construction of the representation $\pi_{\rm mix}$ below provides the fundamental 
building block of canonical quantum field theory for fermions and bosons in the state $\Om$.  
The first step is to introduce an auxiliary canonical algebra $\B$ generated by 
$\{b_i(k),\, b_i^*(k)\, :\, k \in \RR,\, i=1,2\}$, which satisfy 
\bea
&&\left [b_i(k)\, ,\, b_j^*(p)\right ]_{{}_\pm} = 2\pi \delta_{ij}\, \delta(k-p)\II_{\B}\, , 
\label{e7ba}
\\
&&\left [b_i(k)\, ,\, b_j(p)\right ]_{{}_\pm} = \left [b_i^*(k)\, ,\, b_j^*(p)\right ]_{{}_\pm} =0 \, . 
\label{e7bb}
\eea
One can interpret $\{b_i(k),\, b_i^*(k)\}$ 
as annihilation and creation operators of the particle excitations 
at temperature $\beta_i$ and density $\mu_i$ in the reservoir $R_i$. 
In agreement with this idea we adopt the Gibbs representation 
$\{\gamma_{{}_{\rm G}},\, \H_{{}_{\rm G}},\, \Omega_{{}_{\rm G}}\}$
of $\B$. Referring for the details to \cite{Haag, BR}, in the specific case we introduce 
the reservoir Hamiltonians and particle densities 
\begin{equation} 
h_i = \int_{-\infty}^\infty \frac{\rd k}{2\pi} \omega_i (k) b^*_i(k) b_i(k)\, , \qquad 
q_i = \int_{-\infty}^\infty \frac{\rd k}{2\pi} b^*_i(k) b_i(k)\, ,\quad i=1,2 \, .
\label{ah}
\end{equation} 
Following \cite{Haag, BR}, we define 
\begin{equation} 
K = \sum_{i=1}^n \beta_i (h_i -\mu_i q_i) \,  
\label{kop}
\end{equation} 
and introduce the equilibrium Gibbs state over $\B$ in the standard way: for 
any polynomial $\cP$ over $\B$ we set 
\be
\left (\Omega_{{}_{\rm G}},\,  \cP(b_i^*(k_i), b_j(p_j)) \Omega_{{}_{\rm G}} \right ) \equiv 
\langle \cP(b_i^*(k_i), b_j(p_j)) \rangle_{{}_{\rm G}} = 
\frac{1}{Z} \tr \left [\e^{-K} \cP(b_i^*(k_i), b_j(p_j))\right ]\, ,  
\label{def1}
\ee 
where $Z = \tr \left ( \e^{-K}\right )$. It is well known \cite{BR} 
that one can compute the expectation values (\ref{def1}) 
by purely algebraic manipulations using (\ref{e7ba},\ref{e7bb}) and the properties of the trace. 
All these expectation values can be expressed in 
terms of the two-point function 
\begin{equation}
\langle b_i^*(k)b_j(p) \rangle_{{}_{\rm G}} =  
\delta_{ij} 2\pi \delta (k-p)d^\pm_i(k)\, ,  
\label{2a}
\end{equation}
involving the Gibbs distribution (\ref{e10}) of the reservoir $R_i$. 

The final step in the construction of the representation $\pi_{\rm mix}$ 
is to connect the two algebras $\A$ and $\B$ via a specific homomorphism $\gamma$ 
involving the probabilities $w_i(k)$. Using the generators of $\A$ and $\B$, we 
define the mapping $\gamma\, :\, \A \rightarrow \B$ by extending 
\bea 
\gamma \, : \, a(k) &\longmapsto& \sqrt {w_1(k)}\, b_1(k) + \sqrt {w_2(k)}\, b_2(k)
\label{gamma1}
\\
\gamma \, : \, a^*(k) &\longmapsto& \sqrt {w_1(k)}\, b^*_1(k) + \sqrt {w_2(k)}\, b^*_2(k)
\label{gamma2}
\\
\gamma \, : \, \II_{\A} &\longmapsto& \II_{\B} \, , 
\label{gamma3}
\eea
to a homomorphism on the whole $\A$. There is a freedom to insert two independent phase 
factors in (\ref{gamma1},\ref{gamma2}) via the substitution $b_i(k) \mapsto \e^{\ri \sigma_i} b_i(k)$ and 
$b^*_i(k) \mapsto \e^{-\ri \sigma_i} b^*_i(k)$. 
A key property of $\gamma$ is to preserve the canonical structure. Indeed, one has 
\bea 
\left [\gamma[a(k)]\, ,\, \gamma[a^*(p)] \right ]_{{}_\pm} = 
\hskip 4 truecm
\nonumber \\
\left [\sqrt {w_1(k)}\,  b_1(k) + \sqrt {w_2(k)}\, b_2(k)\, ,\, \sqrt {w_1(p)}\, b^*_1(p) + \sqrt {w_2(p)}\, b^*_2(p)\right ]_{{}_\pm} =
\nonumber \\
\left [\sqrt {w_1(k)w_1(p)} +\sqrt {w_2(k)w_2(p)} \right ]2\pi \delta (k-p) \II_{\B} = 2\pi \delta (k-p) \II_{\B}\, , 
\label{can1}
\eea
and analogously for the image of (\ref{e7b}). We observe also that $\gamma$ maps $\A$ to a subalgebra 
$\gamma(\A) \subset \B$ and stress that $\gamma(\A)$ depends on the weights $w_i(k)$. 

At this point, the representation we are looking for, is defined by 
the composition 
\be 
\pi_{\rm mix} = \gamma_{{}_{\rm G}} \circ \gamma \, , 
\label{comp}
\ee
where $\gamma_{{}_{\rm G}}$ is the Gibbs representation of $\B$. 
Applying (\ref{2a}-\ref{gamma3}), one has in fact  
\be 
\langle a^*(k) a(p) \rangle_{\rm mix} = 
w_1(k) \langle b_1^*(k)b_1(p) \rangle_{{}_{\rm G}} 
+ w_2(k) \langle b_2^*(k)b_2(p) \rangle_{{}_{\rm G}} = 
2 \pi \delta (k-p) d^\pm_{\rm mix} (k) \, .
\label{r1}
\ee
We complete the construction of $\pi_{\rm mix}$ by establishing 
the $n$-point functions for $n>2$. In this respect we first observe 
that the particle number 
\be 
{\cal N} = \int_{-\infty}^\infty \rd k\, a^*(k) a(k) \, 
\label{pn1}
\ee 
generates in $\pi_{\rm mix}$ the $U(1)$-phase transformation 
\be 
a(k) \longmapsto \e^{\ri \alpha } a(k)\, , \qquad  a^*(k) \longmapsto \e^{-\ri \alpha } a^*(k)\, , \quad \alpha \in \RR\, .
\label{pn2}
\ee
The conservation of $\cal N$ therefore implies that the only non-vanishing correlation functions of $\{a(k),\, a^*(k)\}$ in 
$\pi_{\rm mix}$ involve equal number of $a^*(k)$ and $a(k)$ operators. A convenient way for representing these 
correlation functions is 
\begin{equation}
\langle a^*(k_1) a(p_1)a^*(k_2) a(p_2)\cdots a^*(k_n) a (p_n)\rangle_{\rm mix} \, ,  \quad n\geq 2\, ,   
\label{pn3}
\end{equation}
where $a^*$ and $a$ are in alternate order. By means of the (anti)commutation relations (\ref{e7a},\ref{e7b}) 
the expectation value with $a^*$ and $a$ in arbitrary order can be always reduced to 
a superposition of $m$-point correlators of the type (\ref{pn3}) with $m\leq n$ . 
The explicit form of (\ref{pn3}) can be expressed in terms 
of the matrix 
\begin{equation} 
\mM^\pm_{ij} = 
\begin{cases} 
2\pi \delta (k_i-p_j)d^\pm_{\rm mix}(k_i)\, ,\qquad \qquad \qquad  i\leq j\, , \\
\mp 2\pi \delta (k_i-p_j)\left [1\mp d^\pm_{\rm mix}(k_i)\right ]\, ,\qquad \; i > j\, .\\ 
\end{cases} 
\label{pn4}
\end{equation} 
where $i,j=1,...,n$. 
In fact, using (\ref{gamma1}-\ref{gamma3}) and the trace formula (\ref{def1}) one can first prove that 
\begin{equation}
\langle a^*(k_1) a(p_1)\cdots a^*(k_n) a (p_n)\rangle_{\rm mix} 
= \sum_{i=1}^n \langle a^*(k_1) a(p_i)\rangle_{\rm mix}\; 
\langle \; \cdots \;  \rangle^\prime_{\rm mix} \, ,
\label{A6bis}
\end{equation} 
where in the primed expectation value $\langle \; \ldots \;  \rangle'_{\rm mix}$ in the right hand side 
the elements $a^*(k_1)$ and $a(p_i)$ have been removed from the string. 
Second, adopting (\ref{A6bis}) one can show by induction in $n$ that 
\begin{equation}
\langle a^*(k_1) a(p_1)\cdots a^*(k_n) a (p_n)\rangle_{\rm mix}   = 
\begin{cases} 
{\rm \bf det}\, [\mM^+]\, , \\ 
{\rm \bf perm}\, [\mM^-]\, ,\\ 
\end{cases} 
\label{pn5}
\end{equation} 
where ${\rm \bf det}$ and ${\rm \bf perm}$ indicate the determinant and the permanent of the corresponding matrices. 
We recall that 
\begin{equation}
{\rm \bf perm}\, [\mM]= \sum_{\sigma_i \in \cP_n} \prod_{i=1}^n \mM_{i \sigma_i} \, , \qquad 
\cP_n - {\rm set\; of\; all\; permutations\; of\; } n\; {\rm elements} \, . 
\label{pn6} 
\end{equation}

It is worth mentioning that the state $\Omega_{\rm mix}$ is not annihilated by the generator $a(k)$. 
For this reason the standard procedure for computing correlation functions in the Fock representation 
is not suitable for our case. The above derivation of (\ref{pn5}) uses instead the homomorphism 
$\gamma$ to map $\A$ to the subalgebra $\gamma (\A) \subset \B$ and adopts \cite{BR} afterwards 
the trace representation (\ref{def1}), which holds in $\B$. 

The above construction has a straightforward generalisation to any number of reservoirs in the Gibbs state. 
The correlation functions in the representation $\pi_{\rm mix}$ 
have still the form (\ref{rep1},\ref{pn5}), where $d^\pm_{\rm mix} (k)$ is the convex combination 
(\ref{mix1}) with $d_i(k)$ given by (\ref{e10}). 

Summarising, we described in this section a general procedure for constructing a 
representation $\pi_{\rm mix}$ of the canonical (anti)commutation relations corresponding to the 
mixture (\ref{mix2}) of the Gibbs distributions associated with 
the heat reservoirs in contact with the system in Fig. \ref{fig1}. In what follows we 
shall investigate a simple example of quantum dynamics illustrating  
the physical properties of $\pi_{\rm mix}$.  
\bigskip 

\section{Physical test of the representation $\pi_{\rm mix}$} 

The conventional description of the system shown in Fig. \ref{fig1} is based on the general Hamiltonian \cite{James-59}
\be 
H_{\rm tot} = H_{\rm bulk} + H_{G_1\cup G_2}+H_{R_1} + H_{R_2} \, . 
\label{ham}
\ee
Here $H_{\rm bulk}$ encodes the time evolution in the bulk $\RR$. $H_{G_1\cup G_2}$ displays  
the interaction in the gates $G_{1,2}$ of the bulk with the reservoirs, which drives the system 
away from equilibrium. Finally, $H_{R_{1,2}}$ describe the reservoirs $R_{1,2}$. 
The main idea of our proposal is to adopt the representation $\pi_{\rm mix}$ for describing the 
system (\ref{ham}). This representation takes into account the interaction 
$H_{G_1\cup G_2}$ by the probabilities $w_{1,2}(k)$ 
and $H_{R_{1,2}}$ by the equilibrium distributions $d_{1,2}^\pm(k)$ of the 
single reservoirs. The scheme allows for any choice of bulk Hamiltonian $H_{\rm bulk}$. 
Since the focus below is on the physical impact of $\pi_{\rm mix}$, 
we will keep the bulk dynamics as simple as possible. 
For this reason we consider a fermionic quantum field $\psi(t,x)$, whose 
time evolution in $\RR$ is fixed by the free Schr\"odinger equation 
\begin{equation}
\ri \prt_t \psi (t,x) +\frac{1}{2m} \prt_x^2 \, \psi (t,x)=0  \, , \qquad m>0 
\label{e1}
\end{equation}
and the equal-time canonical anticommutation relations 
\begin{equation}
[\psi (t,x)\, ,\, \psi^*(t,y)]_{{}_+} = \delta(x-y)\, , \qquad 
[\psi (t,x)\, ,\, \psi (t,y)]_{{}_+} = [\psi^* (t,x)\, ,\, \psi^*(t,y)]_{{}_+} = 0 \, . 
\label{e2}
\end{equation}

The general solution of (\ref{e1},\ref{e2}) is well known and given by 
\be 
\psi (t,x)  =\int_{-\infty }^{\infty} \frac{dk}{2\pi } 
\e^{-\ri \omega (k)t+\ri k x} a (k) \, , 
\quad 
\psi^* (t,x)  =\int_{-\infty }^{\infty} \frac{dk}{2\pi } 
\e^{\ri \omega (k)t-\ri k x} a^* (k) \, , \quad 
\omega(k) \equiv \frac{k^2}{2 m}\, , 
\label{e5}
\ee
where $\{a(k),\, a^*(k)\}$ satisfy the 
anticommutation version of (\ref{e7a},\ref{e7b}) from the previous section. 
In what follows we assume that $\{a(k),\, a^*(k)\}$ in 
(\ref{e5}) are in the representation $\pi_{\rm mix}$. 
After simple algebra one finds the basic two-point functions 
\bea
\langle \psi^*(t_1,x_1) \psi(t_2,x_2)\rangle_{\rm mix} &=& 
\int_{-\infty}^{\infty} \frac{\rd k}{2\pi} \e^{\ri \omega(k) t_{12} -\ri k x_{12}} d^+_{\rm mix} (k)\, , 
\label{corr1} \\ 
\langle \psi(t_1,x_1) \psi^*(t_2,x_2)\rangle_{\rm mix} &=& 
\int_{-\infty}^{\infty} \frac{\rd k}{2\pi} \e^{-\ri \omega(k) t_{12} +\ri k x_{12}} [1-d^+_{\rm mix} (k)]\, . 
\label{corr2}
\eea
Here $t_{12}\equiv t_1-t_2$ and $x_{12}\equiv x_1-x_2$, implying that $\Om$ is invariant under both 
time and space translations. The total energy and momentum are therefore conserved in the system. 
Moreover, (\ref{corr1},\ref{corr2}) are invariant under the phase transformations 
\be 
\psi(t,x) \longmapsto \e^{\ri \alpha } \psi(t,x) \, , \qquad  \psi^*(t,x) \longmapsto \e^{-\ri \alpha } \psi^*(t,x)\, , \quad \alpha \in \RR\, ,
\label{pn2psi}
\ee
induced by (\ref{pn2}). Therefore, in spite of the particle exchange with the heat reservoirs, 
the total particle number of the system is conserved as well. 

One can directly verify that the correlation functions (\ref{corr1},\ref{corr2}) satisfy 
the Kubo-Martin-Schwinger (KMS) condition \cite{Haag, BR} only if $\beta_1=\beta_2$ 
and $\mu_1=\mu_2$. Otherwise the KMS condition is violated, which is a first indication that 
for generic heat bath parameters $\Om$ is a non-equilibrium state. 

A straightforward implication of (\ref{corr1}) concerns the particle density $\varrho (t,x)$ in the state $\Om$. 
Using that $\varrho (t,x) = [\psi^*\psi](t,x)$ one obtains 
\be 
\langle \varrho (t,x)\rangle_{\rm mix} = 
\int_{-\infty}^{\infty} \frac{\rd k}{2\pi} \, d^+_{\rm mix} (k)\, ,  
\label{den1}
\ee
which provides an explicit relation of the mixture $d^+_{\rm mix} (k)$ with a physical observable. 
Because of time and space translation invariance, the mean density does not depend on $t$ and $x$, but 
depends on $\beta_i$, $\mu_i$ and the emission-absorption probabilities $w_i(k)$. The integral 
in (\ref{den1}) is finite and the positivity of $w_i(k)$ and $d_i^+ (k)$ imply the following simple bounds 
\be 
\sum_{i=1}^2{\rm inf}[w_i]\, {\widetilde \varrho} (\beta_i,\mu_i) 
\leq \langle \varrho (t,x)\rangle_{\rm mix} \leq \sum_{i=1}^2{\rm sup}[w_i]\, {\widetilde \varrho} (\beta_i,\mu_i)\, , 
\label{den2}
\ee
with 
\be
{\widetilde \varrho} (\beta,\mu )= -\sqrt {\frac{m}{2\pi \beta}}\, \Li_{1/2}(-\e^{\beta \mu }) \, .
\label{den3}
\ee
Here and in what follows $\Li_s$ stands for the polylogarithm function of order $s$. The overall minus sign in 
(\ref{den3}) compensates the negativity of $\Li_{1/2}(x)$ for $x\leq 0$. 

Using that 
\be 
\lim_{\beta_i \to \infty} d_i^+(k) = \theta (\mu_i -\omega(k)) \, , \qquad i=1,2\, , 
\label{T01}
\ee 
one has in the zero temperature limit 
\be 
\lim_{\beta_1,\, \beta_2 \to \infty} d^+_{\rm mix} (k) = w_1(k) \theta (\mu_1 -\omega(k)) + w_2(k) \theta (\mu_2 -\omega(k))\, , 
\label{T02}
\ee 
which implies in this regime the bounds 
\be 
\frac{1}{\pi} \sum_{i=1}^2 \theta(\mu_i) \sqrt {2 m \mu_i } \, {\rm inf}[w_i] 
\leq \langle \varrho (t,x)\rangle_{\rm mix} \leq  \frac{1}{\pi} \sum_{i=1}^2 \theta(\mu_i) \sqrt {2 m \mu_i } \, {\rm sup}[w_i]\, .  
\label{T03}
\ee

To the end of this section we explore the transport properties of the representation 
$\pi_{\rm mix}$ and the associated entropy production.

\bigskip 

\subsection{Particle transport and full counting statistics}

We investigate here the particle transport in the state $\Om$. For the current operator 
\be
j(t,x)= \frac{\ri }{2m} \left [ \psi^* (\partial_x\psi ) - 
(\partial_x\psi^*)\psi \right ]  (t,x)\, ,   
\label{curr0} 
\ee
one gets by means of (\ref{corr1}) the expectation value  
\be 
\langle j(t,x)\rangle_{\rm mix} = 
\int_{-\infty}^{\infty} \frac{\rd k}{2\pi} \frac{k}{m}\, d^+_{\rm mix} (k)   \, .  
\label{curr1}
\ee
The mean particle flow in $\Om$ is therefore stationary and does 
not depend on the position. We observe moreover that 
(\ref{curr1}) can be equivalently rewritten in the form 
\be 
\langle j(t,x)\rangle_{\rm mix} = 
\int_0^{\infty} \frac{\rd k}{2\pi} \frac{k}{m}I (k) \, [d^+_1 (k)-d^+_2 (k)]\, ,  
\label{curr2}
\ee
which confirms the heuristic argument in section 2.1 that the 
particle exchange imbalance (\ref{Imb4}) is fundamental for the particle flow 
between the reservoirs $R_1$ and $R_2$. In this respect 
we stress that the mean current (\ref{curr2}) vanishes if: 

(a) the two reservoirs are 
at the same temperature and chemical potential, namely
\be 
\beta_1=\beta_2 \equiv \beta\, , \qquad \mu_1 = \mu_2\equiv \mu \, ;  
\label{eq}
\ee 

(b) if the particle emission is fully balanced by the absorption, namely   
\be 
I(k) = 0 \quad \forall \; k \geq 0 \quad \Longleftrightarrow \quad w_i(k) - {\rm even\; function} \, .
\label{I2}
\ee
Such a behaviour is consistent with the fact that in both cases (a) and (b) 
the system is actually in equilibrium, as shown 
later in section 3.4. 

From (\ref{curr2}) it follows that for fixed reservoir distributions 
$d^+_i(\omega(k))$ the profile of $I(k)$ determines both the direction of the current and the energy range 
of the excitations which form the flow. For a more detailed analysis of (\ref{curr2}) it is convenient to 
rewrite it in the form 
\be 
\langle j(t,x)\rangle_{\rm mix} = J(\beta_1,\mu_1) - J(\beta_2,\mu_2) \, , \qquad   
\label{curr2a}
\ee
with 
\be  
J(\beta, \mu) \equiv \int_0^{\infty} \frac{\rd k }{2\pi} \frac{k}{m}I (k) \, d^+(k)\, ,  \qquad   
d^+(k) = \frac{1}{\e^{\beta [\omega (k) - \mu)}+1}
\label{curr2aprime}
\ee
and concentrate on the behaviour of the function $J(\beta,\mu)$.  According to (\ref{T01}), in the  
zero temperature limit one has 
\be 
J(\infty, \mu) = \int_0^{\sqrt {2m\mu}} \frac{\rd k }{2\pi} \frac{k}{m}I (k) \, . 
\label{T04}
\ee  
It is instructive now to compute (\ref{curr2aprime},\ref{T04}) for the imbalances given by (\ref{I3}) and (\ref{I3c}). 
For (\ref{I3}) particles with any energy $\omega (k) \geq 0$ are flowing between the two gates and one finds 
\be 
J(\beta,\mu) = \frac{(c_+ - c_-)}{2\pi \beta} \log \left (1+ \e^{\beta \mu }\right ) \, , \qquad 
J(\infty , \mu) = \frac{(c_+ - c_-)}{2\pi }\, \theta(\mu )\, \mu \, . 
\label{pc1a}
\ee 
With the imbalance (\ref{I3c}) the momenta involved 
in the particle transport are weighted by $\e^{-\xi k^2}$ and one obtains instead 
\be 
J (\beta,\mu) = 
\frac{\e^{\beta \mu }}{2\pi(2m\xi+\beta)}\, \, 
{}_2F_1\left [1,\, 1+ \frac{2m\xi }{\beta},\, 2+ \frac{2m\xi }{\beta},\, -\e^{\beta \mu}\right ]\, , \quad 
J(\infty,\mu) = \frac{1-\e^{-2 m \mu \xi}}{4\pi m \xi}\, , 
\label{pc1b}
\ee
${}_2F_1$ being the hypergeometric function. The impact  of the imbalance 
on the mean particle current (\ref{curr2a}) is evident by comparing (\ref{pc1a}) with (\ref{pc1b}). 

The next step in describing the particle transport in the state $\Om$ is to investigate 
the quantum fluctuations around the mean current (\ref{curr1}), which are given by the connected 
$n$-point functions 
\be 
\langle j(t_1,x_1)\cdots  j(t_n,x_n) \rangle_{\rm mix}^{\rm conn}\, , \qquad n=1,2,...
\label{cs1}
\ee
It is well known that (\ref{cs1}) are the cumulants of a probability distribution $\rho[j]$, which fully describes the 
microscopic features of the quantum transport in the system. Following the fundamental work of Khlus \cite{K-87} 
and Levitov, Lesovik and Chtchelkatchev \cite{L-89}-\cite{LLL-96}, there has been a series of 
contributions deriving $\rho[j]$ for various systems and in different non-equilibrium situations. 
Our goal here will be to establish $\rho[j]$ in the representation $\pi_{\rm mix}$. For this purpose 
we extend to $\Om$ the approach adopted in \cite{MSS-16} for the Landauer-B\"uttiker non-equilibrium 
steady state \cite{L-57}-\cite{B-88}. 

We start by observing that time translation invariance implies that (\ref{cs1}) 
depend actually only on the time differences  
\begin{equation}
\wt_k \equiv t_k - t_{k+1}\, , \quad \, k=1,...,n-1\, , 
\label{td}
\end{equation}
which allows to introduce for $n\geq 2$ the frequency $\nu$ via the Fourier transform 
\be
W^{\rm conn}_n(x_1,...,x_n;\nu ) = 
\int_{-\infty}^{\infty} \rd \wt_1 \cdots   \int_{-\infty}^{\infty} \rd \wt_{n-1} 
\e^{-\ri \nu (\wt_1+\cdots \wt_{n-1})} \langle j(t_1,x_1)\cdots  j(t_n,x_n) \rangle_{\rm mix}^{\rm conn}\, .  
\label{cf8}
\ee 
Following \cite{L-89}-\cite{LLL-96}, we perform the zero-frequency limit 
\begin{equation}
W^{\rm conn}_n[j] = \lim_{\nu \to 0^+} W^{\rm conn}_n (x_1,...,x_n;\nu ) \, .
\label{cf9}
\end{equation} 
In this limit the quantum fluctuations are integrated over 
the whole time axes. It turns out that the position dependence in (\ref{cf9}) drops out 
like in the Landauer-B\"uttiker case \cite{MSS-16}, which represents 
a relevant simplification. 

The main steps in the 
derivation of $W^{\rm conn}_n[j]$ in explicit form are the following: 

(i) by means of (\ref{pn4}-\ref{pn5}) one gets a representation of 
$W^{\rm conn}_n(x_1,...,x_n;\nu )$ involving $n$ integrations over $k_i$ and 
$n$ integrations over $p_j$;

(ii) using the delta functions in (\ref{pn4}) one eliminates all $n$ integrals in $p_j$; 

(iii) plugging the obtained expression in (\ref{cf8}), one performs all $(n-1)$ integrals in $\wt_i$; 

(iv)  at $\nu=0$ the latter produce $(n-1)$ delta-functions, which allow to eliminate the dependence on 
$x_i$ by performing all the integrals over $k_i$ except one, for instance that over $k_1 \equiv k$.   

In this way one finds  
\begin{equation}
W^{\rm conn}_n[j] = \int_{-\infty}^\infty \frac{\rd k }{2\pi} \frac{|k|}{m} \, \C_n (k)\, , 
\label{cs2}
\end{equation} 
where $\C_n (k)$ are the cumulants of the distribution $\rho[j]$ and are compactly represented in the form 
\begin{equation} 
\C_n (k)= (-\ri \partial_\lambda)^n\, \chi (k; \lambda ) \vert_{{}_{\lambda =0}}\, ,  
\label{gft2}
\end{equation}
with the generating function 
\be 
\chi(k; \lambda) = \ln \left [ 1- d^+_{\rm mix} (k) +  d^+_{\rm mix}(k)\, \e^{\ri \lambda \varepsilon (k)} \right ]\, , 
\label{cs3}
\ee
$\varepsilon (k) \equiv \theta(k)-\theta(-k)$ being the sign of $k$. 
The derivation of the the probability distribution $\rho[j]$, we are looking for, is a now simple matter \cite{ST-70}: 
from (\ref{cs3}) one deduces the moment generating function 
\be 
\varphi(k ;\lambda) = \e^{\chi(k ;\lambda)} = 1- d^+_{\rm mix} (k) +  d^+_{\rm mix}(k)\, \e^{\ri \lambda \varepsilon (k)} 
\label{cs5}
\ee 
and performs the Fourier transform
\be 
\rho[j](k ; \xi) = \int_{-\infty}^\infty \frac{\rd \lambda}{2\pi}\, \e^{-\ri \lambda \xi}\, \varphi (k ; \lambda ) \, . 
\label{cs6}
\ee
The result is the Dirac comb function 
\be
\rho[j](k ; \xi) = p_{12}(k ) \delta(\xi -1) + 
p(k )\delta(\xi) + p_{21}(k )\delta(\xi +1)\, ,  
\label{cs7}
\ee
where 
\be 
p_{12}(k) = \theta(k) d^+_{\rm mix} (k)\, ,\qquad 
p(k) = 1-d^+_{\rm mix} (k)\, ,\qquad 
p_{21}(k) = \theta(-k) d^+_{\rm mix} (k)\, ,
\label{cs8}
\ee
satisfy 
\be 
p_{12}(k) + p(k) + p_{21}(k) = 1 \, . 
\label{cs9}
\ee
In addition, using (\ref{w2}, \ref{mix2}) and $0\leq d_i^+(k)\leq 1$ one easily verifies that 
\be 
p_{12}(k),\; p(k),\; p_{21}(k) \geq 0 \, , 
\label{cs91}
\ee
which imply that (\ref{cs7}) is a well defined probability distribution. 

Some comments about the physical interpretation of $\rho[j]$ are in order. 
The arguments of the delta functions in (\ref{cs7}) indicate that 
the probabilities (\ref{cs8}) refer to three different processes in which the particle number changes 
by $\mp 1$ or $0$. Indeed, one can interpret $p_{ij}(k)$ as the probability of a particle to be 
transported from the reservoir $R_i$ to $R_j$, the particle number in the reservoirs changing by 
$\mp 1$.  As far as $p(k)$ is concerned, it can be 
associated with the emission of a particle and its 
subsequent absorption by the same reservoir $R_1$ or $R_2$. In this process  
the particle number of the reservoirs does not change. We observe also that all the probabilities 
(\ref{cs8}) refer to one-particle processes at arbitrary but fixed momentum $k \in \RR$. 
The probabilities for $n$-particle emission and absorption with the same $k$ vanish  
for $n\geq 2$ because of Pauli's principle. This is the reason \cite{MSS-17} why the Dirac comb 
(\ref{cs7}) has only three teeth. For bosonic systems in the Landauer-B\"uttiker steady state 
their number turns out \cite{MSS-18} to be infinite.

\begin{figure}[h]
\begin{center}
\begin{picture}(220,110)(106,25) 
\includegraphics[scale=0.35]{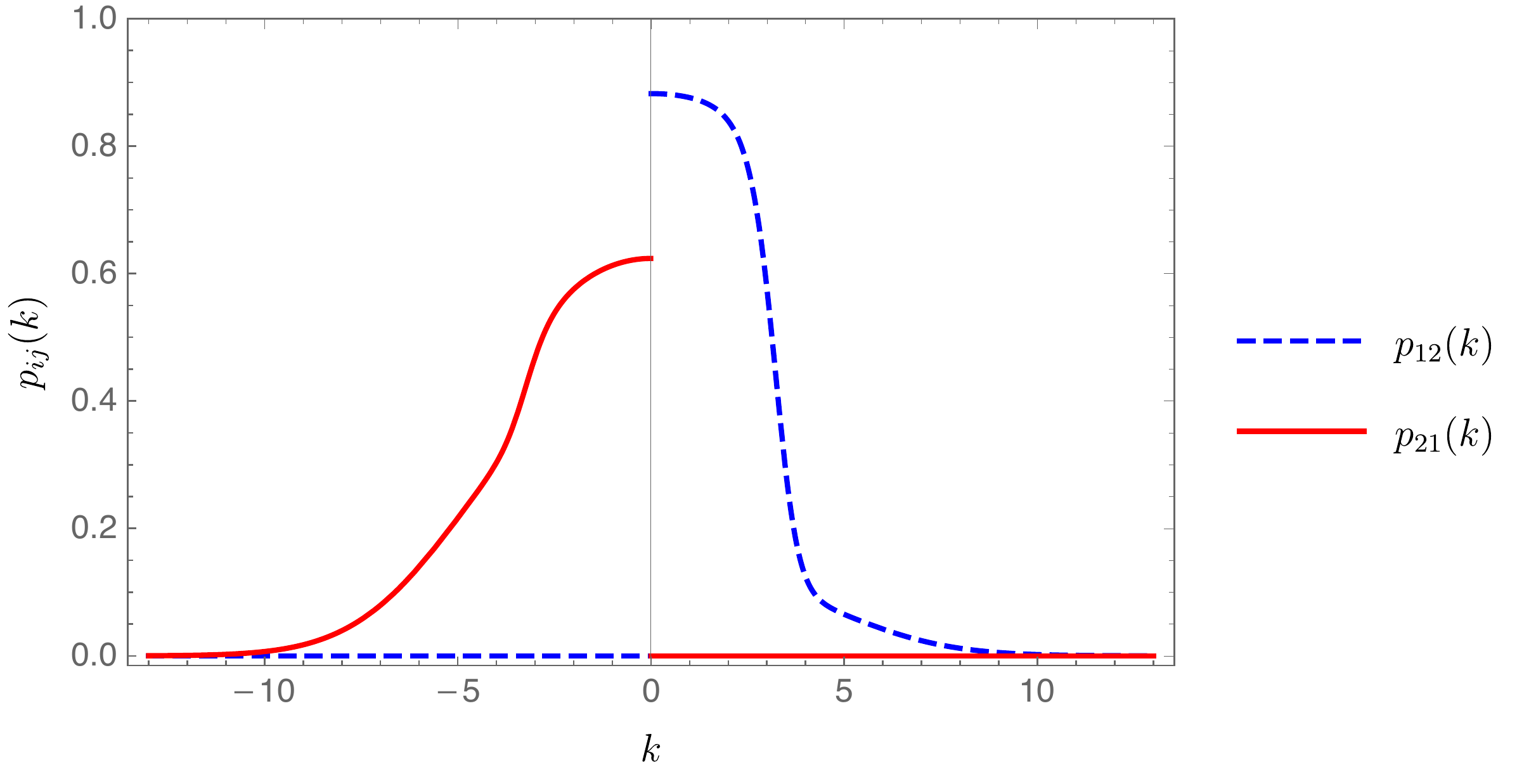}
\hskip 0.5 truecm
\includegraphics[scale=0.35]{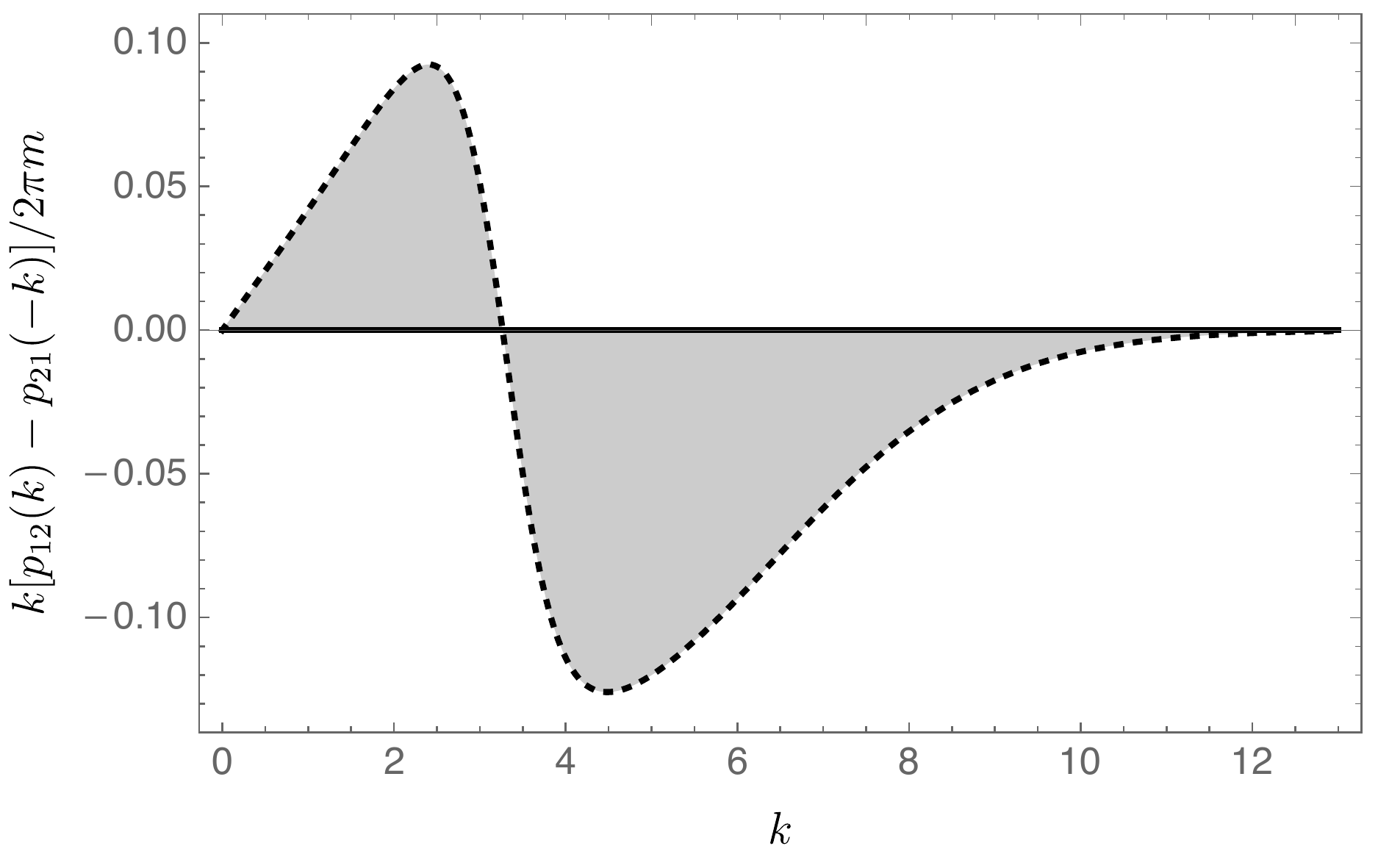}
\end{picture} 
\end{center}
\caption{The probabilities $p_{ij}(k)$ for $c_+=-1/4$, $c_-=1/3$ (left panel) and the integrand of 
equation (\ref{curr21}) (right panel).} 
\label{fig2}
\end{figure}

In view of (\ref{cs9}), one can consider as independent the probabilities 
$\{p_{12}(k),\, p_{21}(k) \}$, which fully characterise 
the particle transport in the state $\Om$ at the microscopic level. 
It is instructive to evaluate $\{p_{12}(k),\, p_{21}(k) \}$ in the cases (a) and (b) 
considered after equation (\ref{curr2}) above. In both of them one gets 
\be 
p_{12}(k) = p_{21}(-k) \not= 0\, , 
\label{cs10}
\ee
which confirms the physical intuition that even at equilibrium the 
quantum fluctuations generate a non-trivial probability of particle exchange between the 
two reservoirs. Nevertheless, the mean current vanishes at equilibrium because (\ref{curr2}) can be 
rewritten in the form 
\be 
\langle j(t,x)\rangle_{\rm mix} = 
\int_0^\infty \frac{\rd k }{2\pi} \frac{k}{m} 
\left [ p_{12}(k)-p_{21}(-k)\right ]\, .  
\label{curr21}
\ee
In order to illustrate (\ref{curr21}) we plotted in the left panel of Fig. \ref{fig2} the probabilities 
$p_{ij}(k)$ for the weights (\ref{w6}) with $c_+=-1/4$ and $c_-=1/3$, fixing the 
heat bath parameters by $\{\beta_1= 0.1,\, \mu_1=2\}$ and $\{\beta_2= 1,\, \mu_2=5\}$ 
in units of mass $m=1$. The area of the shadowed 
region in the right panel gives the mean particle current flowing in the system. 

The probabilities (\ref{cs8}) determine not only the particle transport,  
but provide the common basis for studying the energy and heat transfer as well. They govern 
also the quantum fluctuations of the entropy production explored in section 3.4 below.

\bigskip 

\subsection{Energy and heat flows}

Analogously, one can study the energy transport associated with the current 
\be
\vt (t,x) = \frac{1}{2m} [\left (\partial_t \psi^* \right )\left (\partial_x \psi \right ) 
+ \left (\partial_x \psi^* \right )\left (\partial_t \psi \right ) ](t,x)\, .   
\label{en1}
\ee
The relative expectation value, following from (\ref{corr1}), is 
\bea 
\langle \vt (t,x)\rangle_{\rm mix} = 
\int_{-\infty}^{\infty} \frac{\rd k}{2\pi} \frac{k}{m} \omega(k) d^+_{\rm mix} (k) = 
\qquad \qquad \qquad \quad 
\nonumber \\
\int_0^{\infty} \frac{\rd k}{2\pi} \frac{k}{m}\omega (k) I(k) [d^+_1 (k)-d^+_2 (k)] 
= \int_0^\infty \frac{\rd k }{2\pi} \frac{k}{m} \omega(k) 
\left [ p_{12}(k)-p_{21}(-k)\right ]\, ,  
\label{curr3}
\eea 
which has the same structure like (\ref{curr2},\ref{curr21}), apart of the factor $\omega(k)$ in the integrand. 
The presence of the imbalance $I(k)$ is worth stressing. 
In the case (\ref{I3}) one finds 
\be 
\langle \vt (t,x)\rangle_{\rm mix} = \frac{(c_+ - c_-)}{2\pi} 
\left [\frac{1}{\beta^2_2}\, \Li_2 \left(-\e^{\beta_2\mu_2}\right ) - 
\frac{1}{\beta^2_1}\, \Li_2 \left(-\e^{\beta_1\mu_1}\right )\right ]\, , 
\label{pc2}
\ee 
$\Li_2$ being the dilogarithm. 

\bigskip 

The derivation of the heat flow is more subtle. The point is that the $x$-dependence of the temperature 
and the chemical potential in the bulk $\RR$ of the system are not known away from equilibrium. 
Fortunately, for determining the entropy 
production, it is enough to control the heat currents in the gates $G_i$ where 
$\beta=\beta_i$ and $\mu = \mu_i$. Following \cite{Callen}, for the heat current $q(t, G_i)$ 
in the gate $G_i$ one has   
\begin{equation}
q(t,G_i) = \vt (t,G_i) - \mu_i j(t,G_i)\, .  
\label{h1} 
\end{equation} 
Concerning the currents flowing in the gates $G_i$, we adopt below the following convention. 
With the orientation of the line in Fig. \ref{fig1}, a positive current in the gate $G_1$ is 
incoming in the system, whereas a negative one is outgoing from the system. 
Precisely the opposite holds in the gate $G_2$. Since the particle 
number and the total energy are conserved, the particle and energy inflows are equal to the outflows, namely  
\be
j(t,G_1) = j(t,G_2)\, , \qquad   \vt (t,G_1) = \vt (t,G_2)\, .  
\label{h2} 
\ee 
It is worth stressing the this is not the case for $q(t,G_i)$. In fact, from (\ref{h1}) 
it follows that for $\mu_1\not = \mu_2$ the heat flow in the gate $G_1$ differs from that in $G_2$,
\begin{equation}
\dQ \equiv q(t,G_1)-q(t,G_2) = (\mu_2-\mu_1) j(t,G_1) \not=0\, . 
\label{h3}
\end{equation}
One should recall in this respect that the total energy $E$ of the system has two different components - heat 
and chemical energies. Since $E$ is conserved, (\ref{h3}) implies that these components 
are not separately conserved and can be converted \cite{MSS-15} 
one into the other without dissipation. In our case this general phenomenon is encoded in 
the expectation value 
\begin{equation}
\langle \dQ \rangle_{\rm mix} = (\Om\, ,\, \dQ \Om) \, . 
\label{h4}
\end{equation}
Chemical energy is converted to heat energy if $\langle \dQ \rangle_{\rm mix} < 0$. In the regime 
$\langle \dQ \rangle_{\rm mix} >0$ the system in Fig. \ref{fig1} converts instead heat to chemical energy 
and, as mentioned in section 2, can be compared to a heat engine interpreting the 
chemical as mechanical energy. The efficiency of this engine \cite{qt, he} characterises the quantum transport in the state 
$\Om$ and will be derived below. 
\bigskip

\subsection{Lorenz number and Wiedemann-Franz law}

Employing the above information about the particle and heat transport in the state $\Om$,  
we derive here the Lorenz number in the case (\ref{w6}) and discuss the departure  
from the Wiedemann-Franz (WF) law at finite
temperature. From (\ref{curr2a},\ref{pc1a}) one gets for the fermionic electric 
conductance in the gate $G_i$
\begin{equation}
\sigma (G_i) \equiv e^2 \frac{\der}{\der \mu_i} \langle j(t,x)\rangle_{\rm mix} = 
\frac{(c_+-c_-) e^2}{2\pi (1+ \e^{-\beta_i \mu_i})}\, ,
\label{L1}
\end{equation} 
where the value $e$ of the electric charge has been restored. 
Let us denote by $k_{{}_{\rm B}}$ the Boltzmann constant and let us introduce the temperature 
$T_i = 1/k_{{}_{\rm B}} \beta_i$. Then the thermal conductance in the gate $G_i$ is given by  
\begin{equation}
\kappa(G_i) \equiv \frac{\der}{\der T_i} \langle q (G_i) \rangle  
= -\beta_i^2 k_{{}_{\rm B}} \frac{\der}{\der \beta_i} \langle q (G_i) \rangle \, . 
\label{L2}
\end{equation} 
Combining (\ref{h1}) with (\ref{curr2a},\ref{pc1a},\ref{pc2}), one obtains  
\begin{equation}
\kappa(G_i) = \frac{(c_+-c_-) k_{{}_{\rm B}}}{2\pi \beta_i}
\left \{ \beta_i\mu_i \left [\frac{\beta_i\mu_i}{1+\e^{-\beta_i\mu_i}}-2 \log \left (1+\e^{\beta_i\mu_i} \right ) \right ] 
-\Li_2 \left (-\e^{\beta_i\mu_i}\right ) \right \}\, . 
\label{L3}
\end{equation} 
\begin{figure}[h]
\begin{center}
\begin{picture}(220,122)(20,20) 
\includegraphics[scale=0.43]{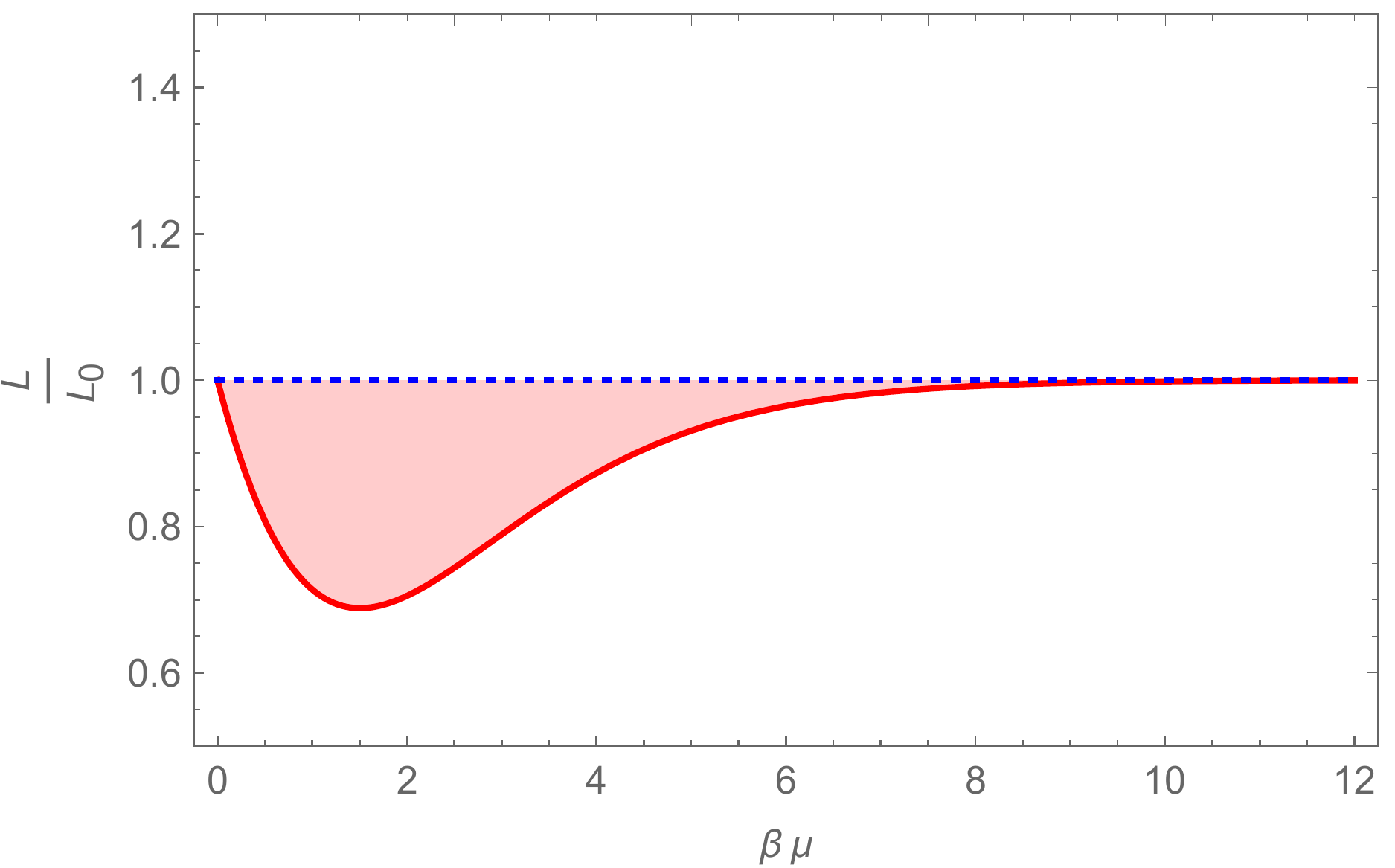}
\end{picture} 
\end{center}
\caption{Shaded zone - finite temperature deviation from the WF law.} 
\label{fig3}
\end{figure} 

The Lorenz number in the gate $G_i$ is defined by the ratio  
\begin{equation}
{\cal L}_i = \frac{\beta_i k_{{}_{\rm B}} \kappa(G_i)}{\sigma (G_i)}\, . 
\label{L4a}
\end{equation}
Employing (\ref{L1}-\ref{L3}) one finds  
\begin{equation}
{\cal L}_i = L(\beta_i \mu_i) \, , 
\label{L4}
\end{equation}
with  
\begin{equation}
L(\beta \mu ) = 
\frac{k^2_{{}_{\rm B}}}{e^2} \left \{ \beta \mu \left [\beta \mu - 2 (1+\e^{-\beta \mu }) \log \left (1+\e^{\beta \mu } \right ) \right ] 
-2 (1+\e^{-\beta \mu })\Li_2 \left (-\e^{\beta \mu }\right ) \right \}\, . 
\label{L5}
\end{equation}
Notice that in our case the Lorenz number ${\cal L}_i$ depends exclusively on the 
dimensionless combination $\beta_i \mu_i$. From (\ref{L5}) one gets 
\be 
\lim_{\beta \mu \to \infty} L(\beta \mu ) = \lim_{\beta \mu \to 0} L(\beta \mu ) = 
\frac{\pi^2 k^2_{{}_{\rm B}}}{3 e^2} \equiv L_0\, ,  
\label{L6}
\ee
which is precisely the value of the free electron Lorenz number. We conclude therefore that 
in the zero and infinite temperature limit the WF law is satisfied in both gates. 
It is satisfied also at any finite temperature for $\mu=0$. 
The deviation from the free electron WF law at finite temperature and $\mu>0$ 
is illustrated in Fig. \ref{fig3}. It is negligible for large values of $\beta \mu$. More precisely, from 
(\ref{L5},\ref{L6})) one gets that $1-L(\beta \mu)/L_0<10^{-3}$ for $\beta \mu \geq 12$. 
Other systems showing deviations from the free electron WF law 
are reported in \cite{KF-97}-\cite{GNESN-22}.

\bigskip 

\subsection{Time reversal breaking and entropy production} 

The entropy production is a fundamental physical quantity characterising the departure from equilibrium. 
It is closely related to the behaviour of the system under time reversal. As is well known \cite{Weinberg},  
the operation of time reversal is implemented by an anti-unitary operator $\mT$. 
In our case $\mT$ acts in the representation $\pi_{\rm mix}$ as follows 
\be 
\mT\, a(k)\, \mT^{-1} = \eta\, a(-k) \, , \qquad \mT\, a^*(k)\, \mT^{-1} = {\overline \eta}\, a^*(-k)\, , \qquad |\eta |^2 = 1 \, . 
\label{NT1}
\ee 
In fact, using (\ref{e5}), (\ref{NT1}) and the anti-unitarity of $\mT$, one easily gets 
\be 
\mT\, \psi(t,x)\, \mT^{-1} = \eta \, \psi(-t,x)\, , \qquad \mT\, \psi^*(t,x)\, \mT^{-1} = {\overline \eta }\, \psi^*(-t,x)\, , 
\label{NT2}
\ee
which, combined with (\ref{curr0}), lead to (see also \cite{Weinberg}) 
\begin{equation}
\mT\, j(t,x)\, \mT^{-1} = -j(-t,x) \, .    
\label{T1}
\end{equation} 

The key point now is the behaviour of the state $\Om \in \H_{\rm mix}$ under time reversal. 
For studying this issue we recall that for a generic configuration with two heat reservoirs
in which (\ref{eq}) and (\ref{I2}) are violated, the mean particle current (\ref{curr0},\ref{curr1},\ref{curr21}) 
in $\Om$ does not vanish, 
\be 
\langle j(t,x)\rangle_{\rm mix} \not= 0\, .
\label{generic}
\ee 
A direct consequence of (\ref{generic}) is that $\mT\, \Om \not = \Om$, namely 
the breaking of time-reversal invariance \cite{MSS-17, MSS-18, BMS-09}. 
This statement can be proven by contradiction. Indeed, the assumption 
$\mT\, \Om = \Om$ leads to  
\begin{equation}
\langle j(t,x) \rangle_{\rm mix}= \langle \mT\, j(t,x)\, \mT^{-1}\rangle_{\rm mix}\, .  
\label{T3}
\end{equation}
On the other hand, the expectation 
value of both sides of (\ref{T1}) gives   
\begin{equation}
\langle \mT\, j(t,x)\, \mT^{-1}\rangle_{\rm mix} = -\langle j(-t,x)\rangle_{\rm mix}\, .  
\label{T4}
\end{equation}
Combining (\ref{T3}) and (\ref{T4}) one gets 
\begin{equation}
\langle j(t,x) \rangle_{\rm mix}= -\langle j(-t,x)\rangle_{\rm mix}\, ,  
\label{T5}
\end{equation}
which is violated because according to equations (\ref{curr0},\ref{curr1},\ref{curr21}) the expectation value 
$\langle j(t,x) \rangle_{\rm mix}$ is time independent. 
This contradiction implies that  
\begin{equation}
\mT\, \Om \not = \Om \, . 
\label{T6}
\end{equation} 

We thus conclude that $\Om$ breaks time-reversal, provided that (\ref{generic}) 
holds and is time independent. This fundamental feature gives 
a strong indication for the presence of a non-trivial entropy production, which is indeed the case. 
In this respect we first observe that with the adopted after equation (\ref{h1}) convention for the incoming and outgoing currents 
in the gates $G_i$ in Fig. \ref{fig1}, 
the entropy production operator is given by \cite{Callen} 
\begin{equation}
\dS(t) = 
\beta_2\, q(t,G_2)  - \beta_1\, q(t,G_1)\, .
\label{ds1}
\end{equation}
Because of (\ref{curr2},\ref{curr3}) the value of the mean entropy 
production in the state $\Om$ is therefore 
\bea 
\langle \dS(t) \rangle_{\rm mix} = \int_{-\infty}^{\infty} \frac{\rd k}{2\pi} 
\frac{k}{m}[\gamma_1(k)-\gamma_2(k)]d^+_{\rm mix} (k) =  
\nonumber \\
\int_0^{\infty} \frac{\rd k}{2\pi} 
\frac{k}{m}\, [\gamma_2(k)-\gamma_1(k)][d^+_1(k) - d^+_2(k)] \, I(k)\, ,  
\label{ds2}
\eea
where 
\be 
\gamma_i(k) \equiv \beta_i [\omega (k) -\mu_i ] \, .
\label{ds3}
\ee
The condition for $\Om$ to be a non-equilibrium state is 
\be 
\langle \dS(t) \rangle_{\rm mix} \not= 0 \, . 
\label{ds3a}
\ee
The stronger condition 
\be 
\langle \dS(t) \rangle_{\rm mix} > 0\, . 
\label{ds4}
\ee
implements in addition the second law of thermodynamics in mean value. 

According to (\ref{ds2}), the constraints (\ref{ds3a}) and (\ref{ds4}) define two integral conditions over the 
positive half line $\RR_+$, which involve both the reservoir distributions and 
the particle exchange imbalance. In order to simplify the analysis of (\ref{ds3a},\ref{ds4}) we observe that 
the two square brackets in the integrand of (\ref{ds2}) have always the same sign because of the identity 
\be 
d_i^+ (k) = \frac{1}{1+ \e^{\gamma_i(k)}}\, . 
\label{e11}
\ee 
Moreover, their product vanishes identically for $k \in \RR_+$ only if both the temperatures and chemical potentials of 
the two heat reservoirs coincide (\ref{eq}). Therefore, excluding this equilibrium case, a sufficient condition for 
(\ref{ds3a}) is that the imbalance has a definite sign in $\RR_+$. A sufficient condition for (\ref{ds4}) is instead 
\be 
I(k) > 0\, , \qquad k\in \RR_+ \, .   
\label{ds5}
\ee 

The above argument shows that the particle exchange imbalance is the key quantity which controls the 
non-equilibrium features of the state $\Om$. For fixed reservoir distributions the 
second law of thermodynamics (\ref{ds4}) generates a relevant physical condition on the 
probabilities $w_i(k)$ which define the mixed distribution. In this respect we observe that the imbalance (\ref{I3c}) 
always satisfies (\ref{ds5}), whereas in the case (\ref{I3}) one must impose 
\be
c_+ >c_-  \, .
\label{ds6}
\ee  

As an application of the inequality (\ref{ds4}), we consider the system in Fig. \ref{fig1} 
in the regime 
\be 
\langle q(t,G_1)\rangle_{\rm mix} >\langle q(t,G_2)\rangle_{\rm mix} >0 \, .
\label{eta0}
\ee
These inequalities imply $\langle \dQ \rangle_{\rm mix} >0$ and therefore, following the discussion at the end 
of section 3.2, the system transforms heat to chemical energy. Accordingly, it can be considered  
as a heat engine \cite{he}, where $R_1$ and $R_2$ are the hot 
and the cold reservoir respectively, namely  
$\beta_1<\beta_2$ (equivalently $T_1>T_2$). 
The mean efficiency is defined by \cite{qt} 
\be 
\eta = \frac{\langle \dQ \rangle_{\rm mix}}{\langle q(t,G_1)\rangle_{\rm mix}}= 
1-\frac{\langle q(t,G_2)\rangle_{\rm mix}}{\langle q(t,G_1)\rangle_{\rm mix}}\, . 
\label{eta1}
\ee
As a consequence of (\ref{ds4}) and (\ref{eta0}) one has 
\be 
\eta <1-\frac{\beta_1}{\beta_2} = 1-\frac{T_2}{T_1}\, , 
\label{eta2}
\ee
reproducing at the quantum level the Carnot bound for the mean efficiency. 

We conclude with some observations about the second law of thermodynamics in the above setting. 
First of all we recall that according to (\ref{ds1}) $\dS(t)$ is an operator 
acting in the state space $\H_{\rm mix}$ of our system. We 
have shown that its mean value in the state $\Om$ is positive, provided that (\ref{ds5}) holds. 
One may wonder if the quantum fluctuation preserve this property as well. We will show now that 
this is indeed the case in the zero frequency limit, i.e. after integrating the fluctuations over the time. 
For this purpose it is convenient to consider the full $n$-point functions 
\be 
W_n[\dS] = \lim_{\nu \to 0^+}
\int_{-\infty}^{\infty} \rd \wt_1 \cdots   \int_{-\infty}^{\infty} \rd \wt_{n-1} 
\e^{-\ri \nu (\wt_1+\cdots \wt_{n-1})}   
\langle \dS(t_1)\cdots  \dS(t_n) \rangle_{\rm mix}\, , \qquad n=2,3,...
\label{css0}
\ee 
In analogy with (\ref{cs2}), the full correlators (\ref{css0}) can be represented in the form 
\be 
W_n[\dS] =  \int_{-\infty}^\infty \frac{\rd k }{2\pi} \frac{|k|}{m} \, \M_n(k)\, , 
\label{css2}
\ee
where now $\M_n(k)$ are the moments of a new probability distribution $\rho[\dS](k ; \sigma)$,  
describing the quantum fluctuations of $\dS(t)$.  
Applying to (\ref{css0}) the procedure (points (i)-(iv) of section 3.1) used for the current correlations, one finds 
\be
\rho[\dS](k ; \sigma) = p_{12}(k) \delta(\sigma -\gamma_{21}(k)) + 
p(k)\delta(\sigma) + p_{21}(k)\delta(\sigma +\gamma_{21}(k))\, .  
\label{css1}
\ee
Here $p_{ij}(k)$ and $p(k)$ are the probabilities given by (\ref{cs8}) 
and 
\be 
\gamma_{ij}(k) \equiv \gamma_i(k)-\gamma_j(k) \, , 
\label{css2bis}
\ee
which can be interpreted as the entropy production generated by the transport of a particle with momentum $k$ 
from the reservoir $R_i$ to $R_j$. Observing that $\gamma_{ij}(k)$ and $\gamma_{ji}(k)$ have opposite sign, 
if the entropy production for the particle transport in the direction $R_i \rightarrow R_j$ is positive, it is negative 
in the direction $R_j \rightarrow R_i$ and vice versa. We already know that the 
probabilities for these two events are $p_{ij}(k)$ and $p_{ji}(k)$ respectively. Therefore, (\ref{cs8}) implies 
that negative entropy production occurs at the quantum level with non-vanishing probability. However, 
it has been shown above that in mean value the positive entropy production 
in the state $\Om$ dominates the negative one if the 
particle exchange imbalance is positive (\ref{ds5}). We will extend now the validity of this statement to 
the zero frequency quantum fluctuations. To this end we consider the moments of (\ref{css1}), 
which are given by 
\be 
\M_n (k) = \int_{-\infty}^\infty \rd \sigma \sigma^n \rho[\dS](k; \sigma) = 
\gamma_{21}^n(k) \left [p_{12}(k) + (-1)^n p_{21}(k) \right ] \, .
\label{css3}
\ee
For even $n=2l$ one has 
\be 
W_{2l}[\dS] =  
\int_{-\infty}^\infty \frac{\rd k }{2\pi} \frac{|k|}{m} \gamma_{21}^{2l}(k) \left [p_{12}(k) + p_{21}(k) \right ] \, ,
\label{css4a}
\ee
which implies 
\be 
W_{2l}[\dS] > 0\, ,
\label{css4b}
\ee
as it should be for the even moments of any probability distribution. 
Using (\ref{cs8}), for $n=2l+1$ on gets instead
\be 
W_{2l+1}[\dS] =  
\int_{-\infty}^\infty \frac{\rd k }{2\pi} \frac{k}{m} \gamma_{21}^{2l+1}(k) d_{\rm mix}^+(k) = 
\int_0^\infty \frac{\rd k }{2\pi} \frac{k}{m} \, \gamma_{21}^{2l+1}(k) \left [d_1^+(k)-d_2^+(k) \right ]I(k) \, ,
\label{css5a}
\ee
which reproduces $\langle \dS(t) \rangle_{\rm mix}$ for $l=0$ and also satisfies 
\be 
W_{2l+1}[\dS] > 0\, ,
\label{css5b}
\ee
provided that (\ref{ds5}) holds. We thus conclude that (\ref{ds5}) implies not only the positivity of the mean 
entropy production, but also of all its zero frequency quantum fluctuations.  
We emphasise that this remarkable property is specific for the entropy production, which  
does not hold for the particle and energy currents. 
For this reason it is tempting to interpret the inequalities 
(\ref{css4b},\ref{css5b}) as a quantum version of the second law of thermodynamics. 
Such interpretation has been already advanced \cite {MSS-17,MSS-18} 
for the Landauer-B\"uttiker non-equilibrium state. 

\bigskip 

\subsection{Quantum noise} 

Once the simple necessary condition (\ref{ds5}) for $\Om$ to be a non-equilibrium state has been established, 
it is instructive to compare the quantum fluctuations away from equilibrium with that at equilibrium. 
We focus here on the noise, namely on the quadratic fluctuations of the particle current (\ref{curr0}) at 
frequency $\nu$. The relative noise power is defined \cite{BB-00} in terms of the connected
two-point current correlation function by the Fourier transform  
\be 
P(\nu; x_1,x_2) = \int_{-\infty}^{\infty} \rd t \, \e^{\ri \nu t} \langle j_x(t,x_1) j_x(0,x_2) \rangle_{\rm mix}^{\rm conn}\, . 
\label{noise1}
\ee
It is well known that $P(\nu;x_1,x_2)$ carries important physical information and provides the basis 
of noise spectroscopy \cite{L-98}. Besides spoiling the particle propagation, 
the noise depends both on the nature of the excitations and the non-equilibrium 
regime in which they are propagated \cite{BB-00}-\cite{MSS-15b}. 
The intriguing issue in our case is the dependence of the noise power (\ref{noise1}) 
on the probabilities $w_{1,2}(k)$, which determine the non-equilibrium features of the state $\Om$. 
Following \cite{BB-00}, we consider the zero frequency limit 
\be 
P_0 = \lim_{\nu \to 0} P(\nu; x_1,x_2) 
=  \int_{-\infty}^{\infty} \frac{\rd k}{2\pi} \frac{|k|}{m} \, \C_2(k)\, ,
\label{noise02}
\ee
where $\C_2(k)$ is the second cumulant of 
the probability distribution $\rho[j](\omega)$. Using (\ref{gft2},\ref{cs3}) one finds  
\be 
P_0 = 
\int_{-\infty}^{\infty} \frac{\rd k}{2\pi} \frac{|k|}{m}\, d^+_{\rm mix} (k) \left [1-d^+_{\rm mix} (k) \right ]\, .
\label{noise2}
\ee

\begin{figure}[h]
\begin{center}
\begin{picture}(220,100)(115,25) 
\includegraphics[scale=0.35]{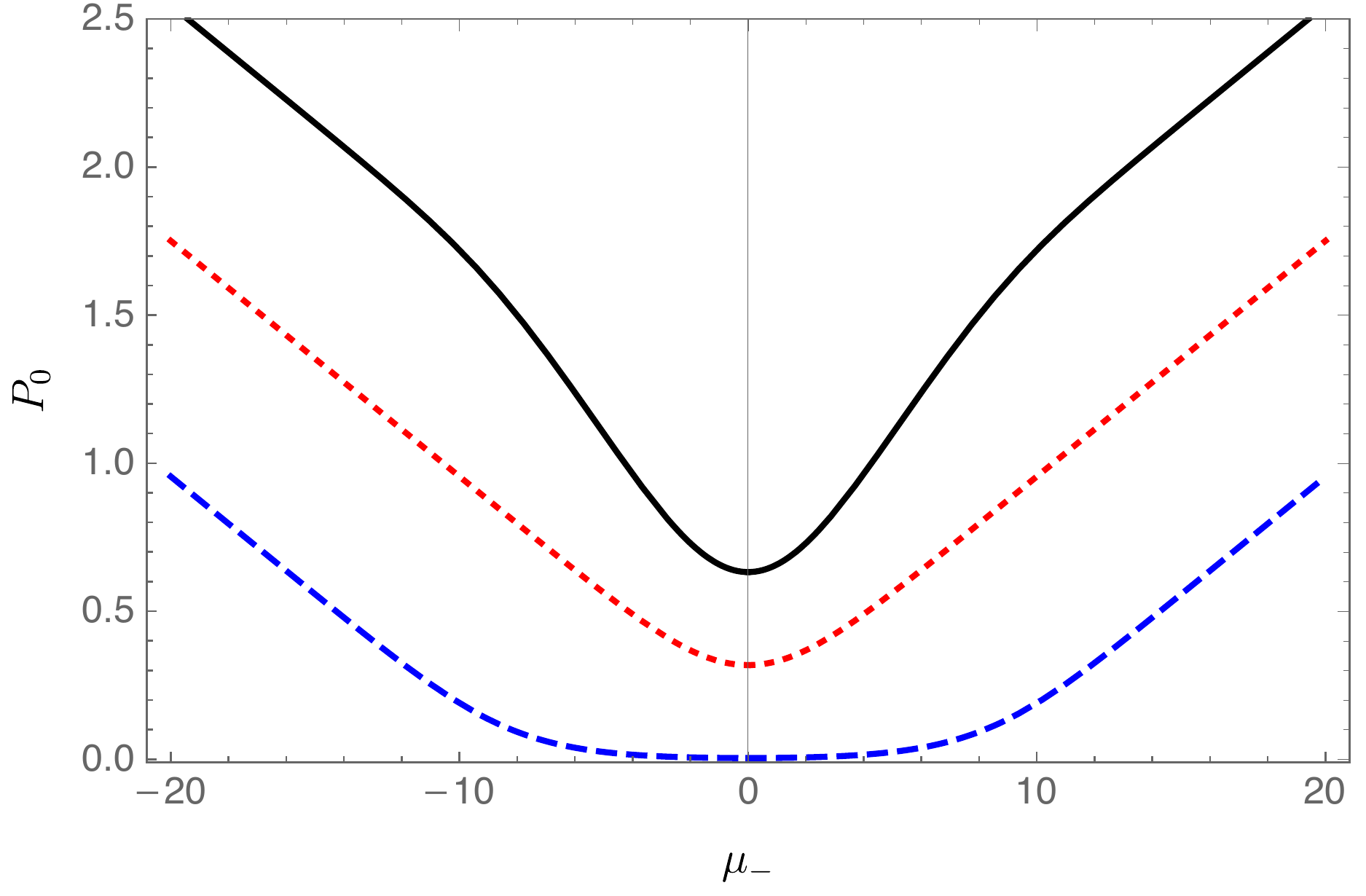}
\includegraphics[scale=0.35]{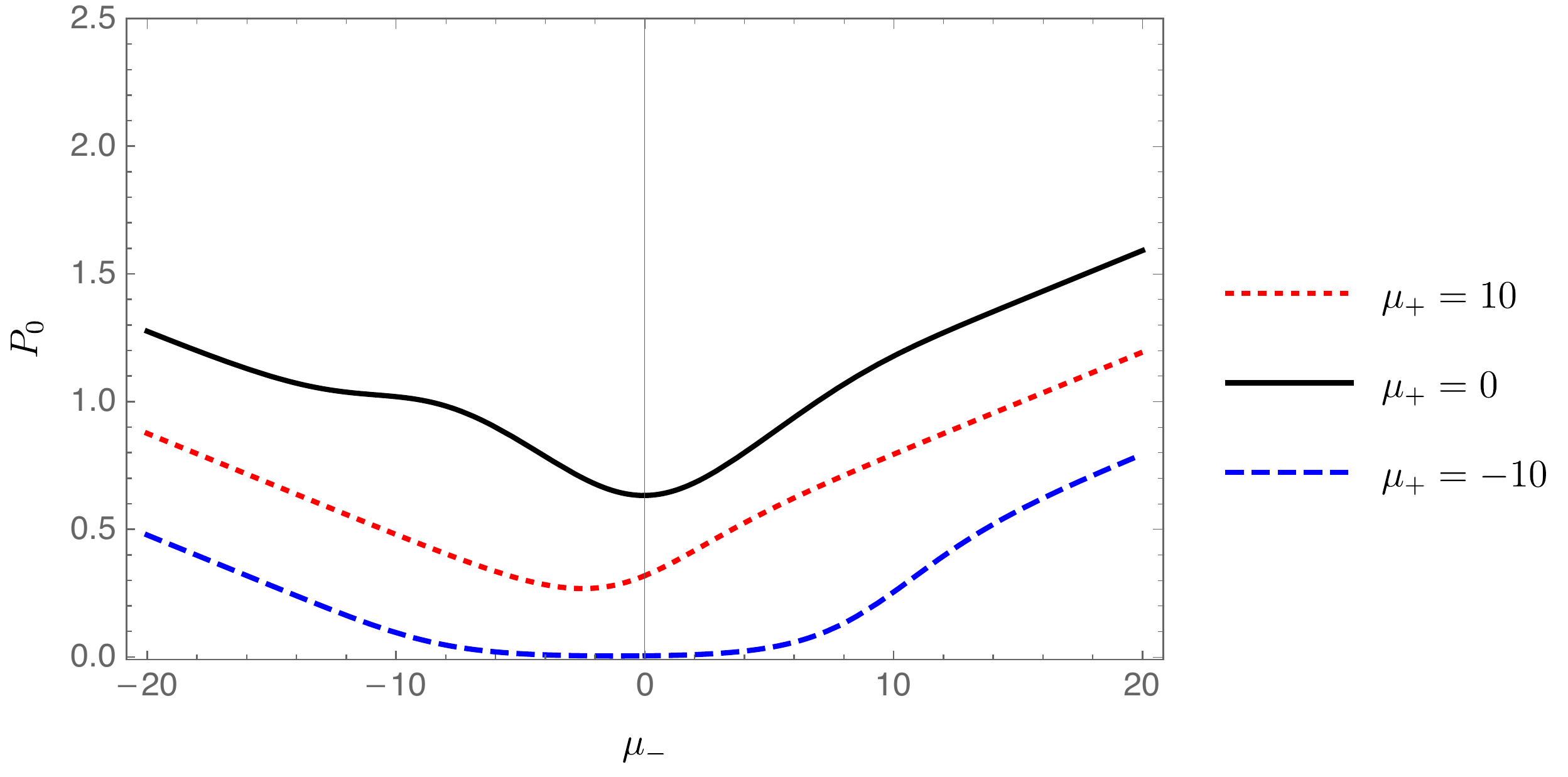}
\end{picture} 
\end{center}
\caption{Zero frequency noise at $\beta_1=\beta_2 =1/2$ for different values of $\mu_+$ at equilibrium $c_+=c_-=0$ (left panel) and away 
from equilibrium $c_+=1/2$, $c_-=0$ (right panel).} 
\label{fig4}
\end{figure}

$P_0$ has a number of remarkable features. First of all 
\be 
P_0 \geq 0 \, , 
\label{positivity}
\ee
which is a consequence of the positivity of the scalar product in the representation $\pi_{\rm mix}$. 
Because of (\ref{w2}), at equilibrium (\ref{eq}) the 
power $P_0$ does not depend on $w_i(k)$ and simplifies to 
\be 
P_0 = 2 \int_0^{\infty} \frac{\rd k}{2\pi} \frac{k}{m} d^+ (k) \left[1 - d^+ (k)\right ] = \frac{1}{\pi \beta (1+\e^{-\beta \mu})} \, ,
\label{noise4}
\ee  
which reproduces at $\mu=0$ the well known Johnson-Nyquist law 
\be 
P_0 = \frac{1}{2\pi \beta} \, .
\label{noise5}
\ee
The dependence of $P_0$ on the interaction with the heat reservoirs via $w_i(k)$ shows up 
away from equilibrium. For illustrating this feature, we specialise (\ref{noise5}) to the case 
(\ref{w6}). One finds 
\bea
P_0 = \int_0^{\infty} \frac{\rd k}{4\pi} \frac{k}{m} \Bigr \{c_{11}\, d_1^+ (k) \left[1- d_1^+ (k)\right ] + 
c_{22}\, d_2^+ (k) \left[1- d_2^+ (k) \right] + 
\nonumber \\
c_{12}\, \left [d_1^+(k) + d_2^+(k) - 2 d_1^+(k)d_2^+(k) \right ]  \Bigr \}\, , \qquad \qquad  
\label{noise6}
\eea
with  
\be 
c_{11}=1+2c_+^2 + 2c_-^2+2c_++2c_-\, , \quad c_{22}=1+2c_+^2 + 2c_-^2-2c_+-2c_- \, , \quad c_{12} = 1-2c_+^2-2c_-^2 \, . 
\ee
For $\beta_1=\beta_2 \equiv \beta$ the integration in (\ref{noise6}) can be performed explicitly and gives 
\be 
P_0 = \frac{1}{4 \pi \beta}\left \{ \frac{c_{11}}{(1+\e^{-\beta \mu_1})} + \frac{c_{22}}{(1+\e^{-\beta \mu_2})} + 
c_{12} \coth \left [ \frac{1}{2} \beta (\mu_1-\mu_2) \right ] \log \left (\frac{1+\e^{\beta \mu_1}}{1+\e^{\beta \mu_2}}\right )\right \}\, . 
\label{noise7}
\ee
In noise experiments \cite{BB-00} $P_0$ is usually expressed in the variables 
\be 
\mu_\pm \equiv \frac{1}{2}(\mu_1-\mu_2)\, . 
\label{noise8}
\ee 
In Fig. \ref{fig4} we use these variables for plotting $P_0$ given by (\ref{noise7}) as a function of the 
difference $\mu_-$ at fixed $\beta$ and $\mu_+$. 
We consider two cases within this setting. The first one, shown in left panel, 
is at equilibrium $\langle \dS \rangle_{\rm mix} = 0$. The right panel displays instead 
the behaviour of the noise power away from equilibrium $\langle \dS \rangle_{\rm mix} > 0$. 
Comparing the two panels, we observe that the departure from equilibrium 
has a specific imprint on $P_0$, which might be useful in experiments.   

\bigskip 

\subsection{Linear response theory in the representation $\pi_{\rm mix}$} 

The above preliminary test of the physical properties of $\pi_{\rm mix}$ is performed 
for freely propagating particles in the bulk. The issue of bulk interactions is 
beyond the scope of this article, but we find it instructive to add a comment about the coupling 
to an external potential $V(x)$ described by 
\begin{equation}
\ri \prt_t \psi (t,x) + \frac{1}{2m} \prt_x^2 \, \psi (t,x) = V(x)\, \psi(t,x)\, . 
\label{int0}
\end{equation}
One can investigate the impact of $V(x)$, adapting the conventional linear response theory (see e.g. \cite{FW}) to the 
non-equilibrium representation $\pi_{\rm mix}$. 
The interaction Hamiltonian generated by the potential $V(x)$ is 
\be 
H_{\rm int}(t) = - \int_{-\infty}^\infty \rd x\, \psi^*(t,x) V(x) \psi(t,x) \, ,
\label{int1}
\ee 
or in momentum space 
\be 
H_{\rm int}(t) = - \int_{-\infty}^\infty \frac{\rd k}{2\pi} \int_{-\infty}^\infty \frac{\rd p}{2\pi}\, a^*(k) {\widehat V}(k-p) a(p) \, , 
\label{int2}
\ee 
where ${\widehat V}(k)$ is the Fourier transform of $V(x)$ and $a^*(k)$ and $a(p)$ are taken in the representation $\pi_{\rm mix}$. 

We focus now on the linear response of the physical observables to the interaction (\ref{int1}). As an example we consider 
the first correction to the mean particle density (\ref{den1}), 
which is given by the second term in the right hand side of 
\be  
\langle \varrho (t,x)\rangle^{\rm lin. resp.}_{\rm mix} = \langle \varrho (t,x) \rangle_{\rm mix} + 
\ri \int_{-\infty}^t \rd \tau \langle [H_{\rm int}(\tau)\, ,\, \varrho(t,x)]_{{}_-}\rangle_{\rm mix} \, .  
\label{lr1}
\ee
Here the novelty with respect to the conventional case \cite{FW} is that the 
four-point functions of $\{a(k)\, ,a^*(k)\}$, appearing in the right hand side of (\ref{lr1}),
are evaluated in the representation $\pi_{\rm mix}$. Using (\ref{int2}), the final result reads  
\bea
\langle \varrho (t,x)\rangle^{\rm lin. resp.}_{\rm mix} = \int_{-\infty}^{\infty} \frac{\rd k}{2\pi}\, d^+_{\rm mix} (k)\ + 
\qquad \qquad \qquad 
\nonumber \\
\int_{-\infty}^{\infty} \frac{\rd k_1}{2\pi} \int_{-\infty}^{\infty} \frac{\rd k_2}{2\pi}\, \frac{{\widehat V}(k_1-k_2)\e^{\ri x(k_1-k_2)}}
{\left [\omega(k_1) - \omega(k_2) - \ri \varepsilon \right]} \left [d^+_{\rm mix} (k_1)+d^+_{\rm mix} (k_2) \right ]\, .  
\label{lr2}
\eea
For the particle current one finds instead, 
\bea
\langle j(t,x)\rangle^{\rm lin. resp.}_{\rm mix} = \int_{-\infty}^{\infty} \frac{\rd k}{2\pi}\, \frac{k}{m} d^+_{\rm mix} (k)\ + 
\qquad \qquad \qquad 
\nonumber \\
\int_{-\infty}^{\infty} \frac{\rd k_1}{2\pi} \int_{-\infty}^{\infty} \frac{\rd k_2}{2\pi}\, \frac{(k_1+k_2){\widehat V}(k_1-k_2)\e^{\ri x(k_1-k_2)}}
{2m \left [\omega(k_1) - \omega(k_2) - \ri \varepsilon \right]} \left [d^+_{\rm mix} (k_1)+d^+_{\rm mix} (k_2) \right ]\, .  
\label{lr3}
\eea
As expected, the correction is linear in the mixed distribution $d^+_{\rm mix} (k)$ and depends 
on the Fourier transform of the potential $V(x)$. Similar expressions are valid for the perturbed 
energy and heat flows.

\bigskip 

\subsection{Main features of the state $\Om$} 

We summarise here the basic physical properties of the state $\Om$:  
\medskip 

(i) $\Om$ is invariant under space and time translations, implying moment and energy conservation in the system; 
\medskip

(ii) the particle number in the state $\Om$ is conserved and the mean particle density does not vanish; 
\medskip 

(iii) the mean values of the particle and energy currents in $\Om$ define 
steady flows, which depend on the particle exchange imbalance $I(k)$; 
\medskip

(iv) $\Om$ breaks down time reversal invariance and gives rise to non-trivial 
entropy production $\langle \dS \rangle_{\rm mix}$;
\medskip 

(v) the condition $I(k) > 0$ implies that the mean value of the entropy production operator $\dS(t)$ 
and all its zero frequency quantum fluctuations in the state $\Om$ are positive;  

\medskip 

(vi) the zero-frequency noise in the state $\Om$ carries a specific dependence on 
the probabilities $w_i(k)$ and is sensitive to the departure from equilibrium;
\medskip

(vii) conventional linear response theory works also in the representation $\pi_{\rm mix}$. 
\medskip 

The above properties have a universal character and 
indicate that the representation $\pi_{\rm mix}$ has solid physical grounds. 
On this basis we expect that it can be successfully applied for describing the 
non-equilibrium dynamics of a variety of other quantum systems in contact with two heat reservoirs.

\bigskip

\section{Discussion and outlook}

The fundamental role of the canonical (anti)commutation relation algebra $\A$ and its equilibrium representations 
(Fock, Gibbs,...) is well known. The main achievement in this paper 
is the explicit construction of a non-equilibrium representation $\pi_{\rm mix}$ of $\A$, inspired by 
the physics of quantum systems in contact 
with two or more heat reservoirs. The construction is based on the concept of mixture 
of probability distributions (\ref{mix1}) and has two main ingredients: the equilibrium distributions 
$\{d_i(k) \, :\, k \in \RR, \, i=1,...,N \}$ describing 
the heat reservoirs and the probabilities (weights) $\{w_i(k) \, :\, k \in \RR, \, i=1,...,N \}$ of 
particle exchange of each reservoir with the system. 
In this way the physics of the heat reservoirs and their contact with the system is fully encoded in 
$\pi_{\rm mix}$ and one is left with the choice of bulk dynamics. In this respect the framework 
is very general and applies without restrictions to any type of quantum evolution in the bulk, 
which is subject to the equal time canonical (anti)commutation relations. The construction 
of both non-relativistic and relativistic quantum fields in the representation $\pi_{\rm mix}$ is straightforward. 

We recall that our goal in this paper is to construct and investigate steady states \cite{James-59}-\cite{Kita-10} 
for the system in Fig. \ref{fig1}. 
As already mentioned in the introduction, for this purpose we consider reservoirs with large enough capacity, 
so that the emission and absorption of particles do not alter their temperature and chemical potential. 
Relaxing this assumption, one can introduce \cite{SNB-14}-\cite{ABW-20} time-dependent 
parameters $\beta_i(t)$ and $\mu_i(t)$, which take into account the evolution of the reservoir parameters, 
but lead to non-stationary states. 

In order to get a first general idea about the physical properties of the representation $\pi_{\rm mix}$, we 
explored the free Schr\"odinger quantum field theory in the bulk. In this context we explicitly 
evaluated various physical observables, including 
the particle density in the state $\Om$ and the mean particle and 
heat currents. The probability distribution, governing the 
zero frequency quantum fluctuations of the particle current (full counting statistics), was 
established as well. We also explored the Lorenz number, the entropy production 
and the noise power in the system. All of them exhibit 
meaningful and realistic physical behaviour. It will be very interesting to test the representation 
$\pi_{\rm mix}$ in other continuous or lattice systems with non-trivial bulk interactions. 
An intriguing possibility in this respect is the implementation of the mixed distributions 
(\ref{mix1},\ref{mix2}) in the setup of non-equilibrium conformal field theory 
\cite{MS-13,BD-15,HL-18,MS-20}. 

We discussed in detail the mixed distribution (\ref{mix2}) corresponding to 
$N=2$ heat reservoirs described by Gibbs distributions. 
Our framework however can be easily extended to $N>2$ reservoirs $R_i$ in the Gibbs or 
other equilibrium distributions like the generalised Gibbs ensemble \cite{J1,J2}. 
A typical example is a space with the geometry of a star-graph with vertex $V$ and 
$N$ edges. The reservoirs are attached at the end-points of the edges. 
Such geometry is employed in the literature \cite{kf-92}-\cite{KGA-20} as a model of 
multilead quantum wire junction. 
Defining the particle exchange imbalance in the gate $G_i$ by (\ref{Imb1}) with $i=1,2,...,N$, 
as a consequence of (\ref{w1}) one has 
\be 
\sum_{i=1}^N I_i(k) = 0\, , 
\label{ng2}
\ee
which is the generalisation of (\ref{Imb3}). The number of independent imbalances, which characterise 
the departure from equilibrium in this case are therefore $N-1$. 

Concerning other physical applications of the representation $\pi_{\rm mix}$, 
we recall that there exist several non-equilibrium 
processes like dissipation, diffusion, relaxation and decoherence. 
They are closely related \cite{CH-08} and describe different features of quantum irreversibility. 
The remarkable progress in out of equilibrium field theory in the last decade 
reveals (see the recent reviews \cite{RV-21,ABF-21} and references therein) 
that these processes have a deep impact on various aspects of many-body 
quantum physics including critical phenomena, phase transitions, symmetry breaking, 
quench dynamics and entanglement. A challenging open problem is to explore in this wider context the physics 
described by the non-equilibrium representation $\pi_{\rm mix}$. Last but not least is the extension 
of the framework to higher space dimensions.

\bigskip 
\bigskip
\leftline{\bf Acknowledgments} 
\medskip 

It is a great pleasure to thank my friend and long-time collaborator Paul Sorba for the constant interest in 
the subject of this investigation. Insightful discussions with Erik Tonni and Ettore Vicari are also kindly acknowledged.

\end{document}